\documentclass[fleqn,usenatbib,useAMS]{mnras}

\usepackage[T1]{fontenc}
\usepackage{ae,aecompl}
\usepackage{graphicx}	
\usepackage{amsmath}	
\usepackage{amssymb}	
\usepackage{xspace}
\usepackage[table]{xcolor}

\usepackage{scalerel}
\usepackage{tikz}
\usetikzlibrary{svg.path}
\definecolor{orcidlogocol}{HTML}{A6CE39}
\tikzset{
  orcidlogo/.pic={
    \fill[orcidlogocol] svg{M256,128c0,70.7-57.3,128-128,128C57.3,256,0,198.7,0,128C0,57.3,57.3,0,128,0C198.7,0,256,57.3,256,128z};
    \fill[white] svg{M86.3,186.2H70.9V79.1h15.4v48.4V186.2z}
                 svg{M108.9,79.1h41.6c39.6,0,57,28.3,57,53.6c0,27.5-21.5,53.6-56.8,53.6h-41.8V79.1z M124.3,172.4h24.5c34.9,0,42.9-26.5,42.9-39.7c0-21.5-13.7-39.7-43.7-39.7h-23.7V172.4z}
                 svg{M88.7,56.8c0,5.5-4.5,10.1-10.1,10.1c-5.6,0-10.1-4.6-10.1-10.1c0-5.6,4.5-10.1,10.1-10.1C84.2,46.7,88.7,51.3,88.7,56.8z};
  }
}

\usepackage{newtxtext,newtxmath}

\newcommand\orcidicon[1]{\href{https://orcid.org/#1}{\mbox{\scalerel*{
\begin{tikzpicture}[yscale=-1,transform shape]
\pic{orcidlogo};
\end{tikzpicture}
}{|}}}}

\newcommand{\angstrom}{\textup{\AA}}

\newcommand{\mspeak}{\ensuremath{M_{*, \mathrm{peak}}}\xspace}

\newcommand{\Mstar}{\ensuremath{M_*}\xspace}

\newcommand{\Msun}{\ensuremath{\mathrm{M}_\odot}\xspace}

\newcommand{\Msunyr}{\ensuremath{\mathrm{M}_\odot\,\mathrm{yr}^{-1}}\xspace}

\definecolor{hpurple}{HTML}{7E16DF}

\newcommand{\logMpeak}{\ensuremath{\log_{10} (M_{\rm peak})}\xspace}

\newcommand{\sigmasfr}{\ensuremath{\sigma(\Delta_\mathrm{SFR})}\xspace}
\newcommand{\sigmacol}{\ensuremath{\sigma(\Delta_c)}\xspace}

\newcommand{\fsps}{\textsc{fsps}\xspace}

\newcommand{\um}{\textsc{universemachine}\xspace}
\newcommand{\tng}{IllustrisTNG\xspace}

\definecolor{Gray}{gray}{0.9}
\newcolumntype{a}{>{\columncolor{Gray}}c}

\newcommand{\bit}{\begin{enumerate}}
\newcommand{\eit}{\end{enumerate}}

\title[On forward modelling galaxy colours]{Surrogate modelling the Baryonic Universe II: on forward modelling the colours of individual and populations of galaxies}

\author[Chaves-Montero \& Hearin]{
Jon\'{a}s Chaves-Montero$^{1,2}$\thanks{E-mail: \href{mailto:jonas.chaves@dipc.org}{jonas.chaves@dipc.org}}\orcidicon{0000-0002-9553-4261} and Andrew Hearin$^{2}$\orcidicon{0000-0003-2219-6852}
\\$^{1}$ Donostia International Physics Centre, Paseo Manuel de Lardizabal 4, 20018 Donostia-San Sebastian, Spain.
\\$^{2}$ HEP Division, Argonne National Laboratory, 9700 South Cass Avenue, Lemont, IL 60439, USA.
}

\date{}
\pubyear{2021}

\begin{document}
\label{firstpage}
\pagerange{\pageref{firstpage}--\pageref{lastpage}}
\maketitle

\begin{abstract}
Among the properties shaping the light of a galaxy, the star formation history (SFH) is one of the most challenging to model due to the variety of correlated physical processes regulating star formation. In this work, we leverage the stellar population synthesis model \fsps, together with SFHs predicted by the hydrodynamical simulation \tng and the empirical model \um, to study the impact of star formation variability on galaxy colours. We start by introducing a model-independent metric to quantify the {\em burstiness} of a galaxy formation model, and we use this metric to demonstrate that \um predicts SFHs with more burstiness relative to \tng. Using this metric and principal component analysis, we construct families of SFH models with adjustable variability, and we show that the precision of broad-band optical and near-infrared colours degrades as the level of unresolved short-term variability increases. We use the same technique to demonstrate that variability in metallicity and dust attenuation presents a practically negligible impact on colours relative to star formation variability. We additionally provide a model-independent fitting function capturing how the level of unresolved star formation variability translates into imprecision in predictions for galaxy colours; our fitting function can be used to determine the minimal SFH model that reproduces colours with some target precision. Finally, we show that modelling the colours of individual galaxies with percent-level precision demands resorting to complex SFH models, while producing precise colours for galaxy {\em populations} can be achieved using models with just a few degrees of freedom.
\end{abstract}



\section{Introduction}

The modelling of galaxy spectral energy distributions (SEDs) has wide-ranging applications in astronomy, from gaining insight into the fundamental properties that shape galaxy evolution to setting unbiased constraints on cosmological parameters. The most common framework used to predict SEDs is stellar population synthesis \citep[SPS, for a review see][]{Conroy2013}, which relies on stellar evolution theory to model galactic light from the ultraviolet to the infrared. The use of SPS to model galaxy SEDs dates back over forty years \citep{Tinsley1978_sps, Bruzual1983_sps, Arimoto1987_sps, Guiderdoni1987_sps, Buzzoni1989_sps}, and this remains an active field of research today at least in part because of the multitude of correlated physical processes that influence the SED, including star formation history (SFH) \citep{heavens_etal00}, metallicity \citep{tojeiro_etal07}, and extinction by dust \citep{conroy_etal10a}.

Among all the features of a galaxy that influence its colours, the star formation history is one of the most challenging to model due to the combination of two effects: galaxy formation models predict a huge diversity of complex SFHs \citep{Sparre2015, Matthee_Schaye2019, Iyer2020, joshi2021_CumulativeStarformationhistories} and the information encoded in galaxy's emitted light about SFHs is very limited even for high-resolution optical spectra \citep{Ocvirk2006}. As a result, SPS models typically describe SFHs using either functional forms controlled by a small number of free parameters, or alternatively, by star formation rates (SFRs) defined at a reduced set of cosmic times. Some examples of typical parametric models are exponentially declining \citep{Schmidt1959}, delayed exponential \citep{Sandage1986_sfhs}, increasing \citep{Buat2008, Lee2009}, lognormal \citep{Gladders2013, diemer_etal17}, and two-epoch SFHs \citep{Behroozi2013, simha_etal14, Ciesla2017, tinker17, Carnall2018}. The main advantage of this type of models is that relatively simple functional forms capture most SFHs predicted by simulations with reasonable precision \citep{simha_etal14, diemer_etal17}, although it should be noted that substantial biases on the inferred galaxy properties are still possible in many cases \citep{michalowski2012_StellarMassesspecific, leja_etal19, Lower2020}. On the other hand, non-parametric models predict a wider variety of galaxy features but at a higher computational cost; some examples of these models are piecewise constant \citep{CidFernandes2005, Ocvirk2006}, adaptive \citep{tojeiro_etal07}, stochastically correlated \citep{caplar_tachella19, tacchella2020_StochasticModellingstarformation}, Gaussian Processes based \citep{iyer_etal19}, and principal component based \citep{Sparre2015, Matthee_Schaye2019, Chen2020_empiricalSFH} SFHs. We refer the reader to \citet{Carnall2019} and \citet{leja_etal19} for a recent comparison between parametric and non-parametric methods.

Although short-term star formation fluctuations have been detected in observations \citep{Weisz2012_variability, guo2016_BURSTYSTARFORMATION, caplar_tachella19, broussard2019_StarFormationStochasticity, wang2020_VariabilityStarFormationa, Wang_Lilly_2020} and are a ubiquitous prediction of galaxy formation models \citep{Sparre2017, Iyer2020}, such features are not captured by many of the widely-used SFH models considered by SPS codes. Motivated by this, we study how time variability in galaxy properties affects the synthesis of broad-band optical and near-infrared colours. We isolate the impact of variability on galaxy colours following a similar approach as in \citep[][hereafter CH20]{chaves_hearin2020_sbu1}: we produce colours using the SPS model \fsps \citep{Conroy2009, conroy_gunn_2020} while holding fixed all galaxy properties besides that studied. We start by generating colours using SFHs predicted by the cosmological hydrodynamical simulation \tng \citep{Pillepich2018a} and the empirical model \um \citep{behroozi_etal19}, and we study how these colours compare to those arising from approximate SFH models that capture adjustable levels of variability. Then, we address the influence on galaxy colours of variability in metallicity and dust attenuation, and we compare the relative impact of variability on these properties to that on star formation.

This work is the second in a series building a new approach to simulation-based forward-modelling the galaxy-halo connection: Surrogate modelling the Baryonic Universe (SBU). In the first paper of the series (CH20), we studied the influence of star formation history on galaxy colours, and identified a unique direction in colour-colour space that corresponds to the most common variations amongst physically-motivated SFHs. In the present work, we develop a formalism for studying physically-motivated variability in galaxy properties. We use our methodology to quantify the ability of an approximate SFH model to resolve short-term fluctuations, and we study how the presence of unresolved star formation fluctuations impacts model predictions for broad-band photometry. One of the main conclusions of this work is that even though the impact of unresolved fluctuations in galaxy properties on the colours of individual galaxies is substantial, the distributions of colours of a galaxy population are relatively insensitive to short-term variability. Taking advantage of this important simplification, in a companion paper to the present work \citep{Hearin2021} we build a computationally efficient, simulation-based surrogate model for the prediction of broad-band galaxy colours.

We do not attempt to study the impact of variability on the synthesis of galaxy colours at a particular redshift; we instead address the effect of variability on the entire observed-frame colour history. This distinction is critical because, for example, galaxy colours produced at a specific redshift depend primarily on the relative amount of stellar mass formed over the last billion years before that redshift (CH20). Consequently, star formation fluctuations taking place a few million years before study time present a much more substantial impact on colours than fluctuations happening 3 billion years before such time. Studying the impact of variability on observed-frame colour histories is motivated by the goals of SBU: we will forward-model galaxy colours as a function of the assembly history of their host haloes, in a similar fashion as semi-analytic models \citep[e.g.,][]{kauffmann_white_Guiderdoni_1993, baugh_cole_frenk_1996, avila_reese_1998}. The primary purpose of this work is to quantify the precision of galaxy colours synthesised using star formation, metallicity, and dust attenuation histories as a function of the level of variability resolved by these.

Forward-modelling galaxy catalogues is essential for ongoing experiments such as the Dark Energy Survey\footnote{\url{https://www.darkenergysurvey.org/}} \citep[DES,][]{des}, the Hyper-Suprime Cam\footnote{\url{https://hsc.mtk.nao.ac.jp/ssp/}} \citep[HSC,][]{hsc}, and the Kilo-Degree Survey\footnote{\url{http://kids.strw.leidenuniv.nl/}} \citep[KiDS,][]{kids}, and near-future surveys like the Javalambre Physics of the Accelerating Universe Astrophysical Survey\footnote{\url{http://www.j-pas.org/}} \citep[J-PAS,][]{jpas}, the Rubin Observatory Legacy Survey of Space and Time\footnote{\url{http://www.lsst.org/}} \citep[LSST,][]{ivezic_etal08, lsst_science_book}, Euclid\footnote{\url{http://www.euclid-ec.org/}} \citep{laureijs11}, the Spectro-Photometer for the History of the Universe, Epoch of Reionization and Ices Explorer\footnote{\url{http://spherex.caltech.edu/}} \citep[SPHEREx,][]{spherex}, and the Wide Field Infrared Survey Telescope \citep[WFIRST,][]{wfirst}. The majority of galaxies observed by these surveys do not present redshift estimates precise enough to translate observed-frame to rest-frame colours, explaining our decision to analyse observed-frame colours in this work. Our aim is to apply the SBU approach to mimic the measurements of these and other surveys, thereby enabling a precise characterisation of the impact of systematic uncertainties on cosmological parameters, as well as setting constraints on galaxy formation physics using large-scale structure data.

The paper is organised as follows. In \S\ref{sec:data}, we describe the main properties of the galaxy formation models from which we draw star formation histories: the cosmological hydrodynamical simulation \tng and the empirical model \um. Then, in \S\ref{sec:model}, we use principal component analysis to develop SFH models presenting adjustable levels of star formation variability, which we characterise in detail. In \S\ref{sec:colours}, we use these techniques to study the impact of unresolved star formation fluctuations on galaxy colours. In \S\ref{sec:forward}, we study the performance of different SFH models predicting the colours of individual and populations of galaxies. Finally, we discuss the consequences of our results for forward-modelling galaxy colours in \S\ref{sec:discussion}, and in \S\ref{sec:conclusions} we summarise our principal findings.

Throughout this paper we use {\it Planck} 2\,015 cosmological parameters \citep{planck14b}: $\Omega_{\rm m}= 0.314$, $\Omega_\Lambda = 0.686$, $\Omega_{\rm b} = 0.049$, $\sigma_8 = 0.83$, $h = 0.67$, and $n_{\rm s} = 0.96$. Magnitudes are in the AB system and simulated colours are computed by convolving \fsps spectra with LSST transmission curves\footnote{\url{https://github.com/lsst/throughputs/tree/master/baseline/}}. Halo and stellar masses are in $h^{-1}M_{\odot}$ and $M_{\odot}$ units, respectively.


\section{Datasets}
\label{sec:data}

In order to study the impact of short-term star formation fluctuations on colours, for our validation data we use SFHs predicted by contemporary galaxy formation models. The main drawback of this approach is that our results have a degree of model dependence; to mitigate this, we draw physically-motivated SFHs from two different galaxy formation models built upon entirely distinct principles: the cosmological hydrodynamical simulation \tng \citep{Marinacci2018, Naiman2018, Nelson2018b, Pillepich2018a, Springel2018} and the empirical model \um \citep{behroozi_etal19}. In what follows, we briefly describe the main features of these models.


\begin{figure*}
    \centering
    \includegraphics[width=\columnwidth]{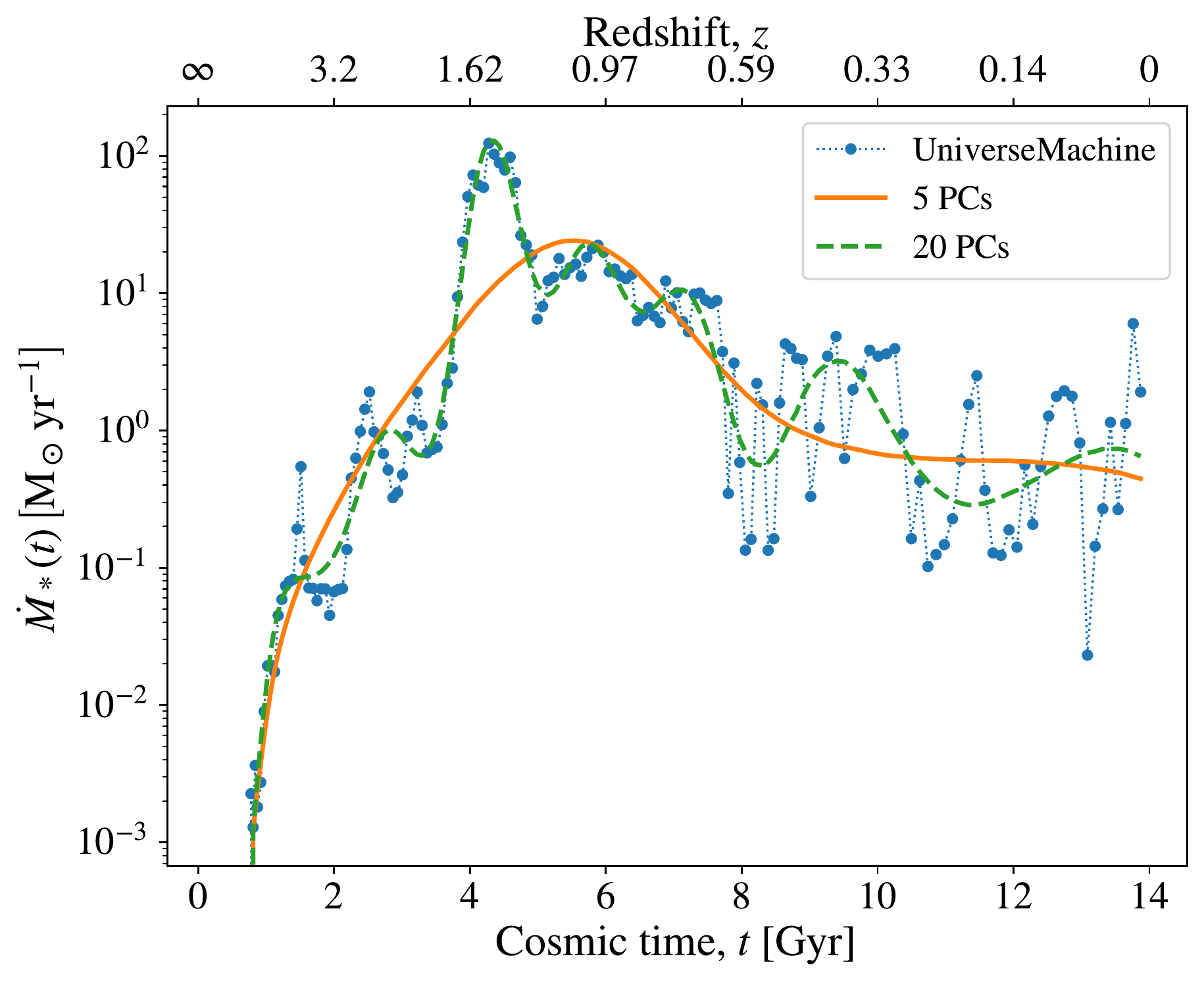}
    \includegraphics[width=\columnwidth]{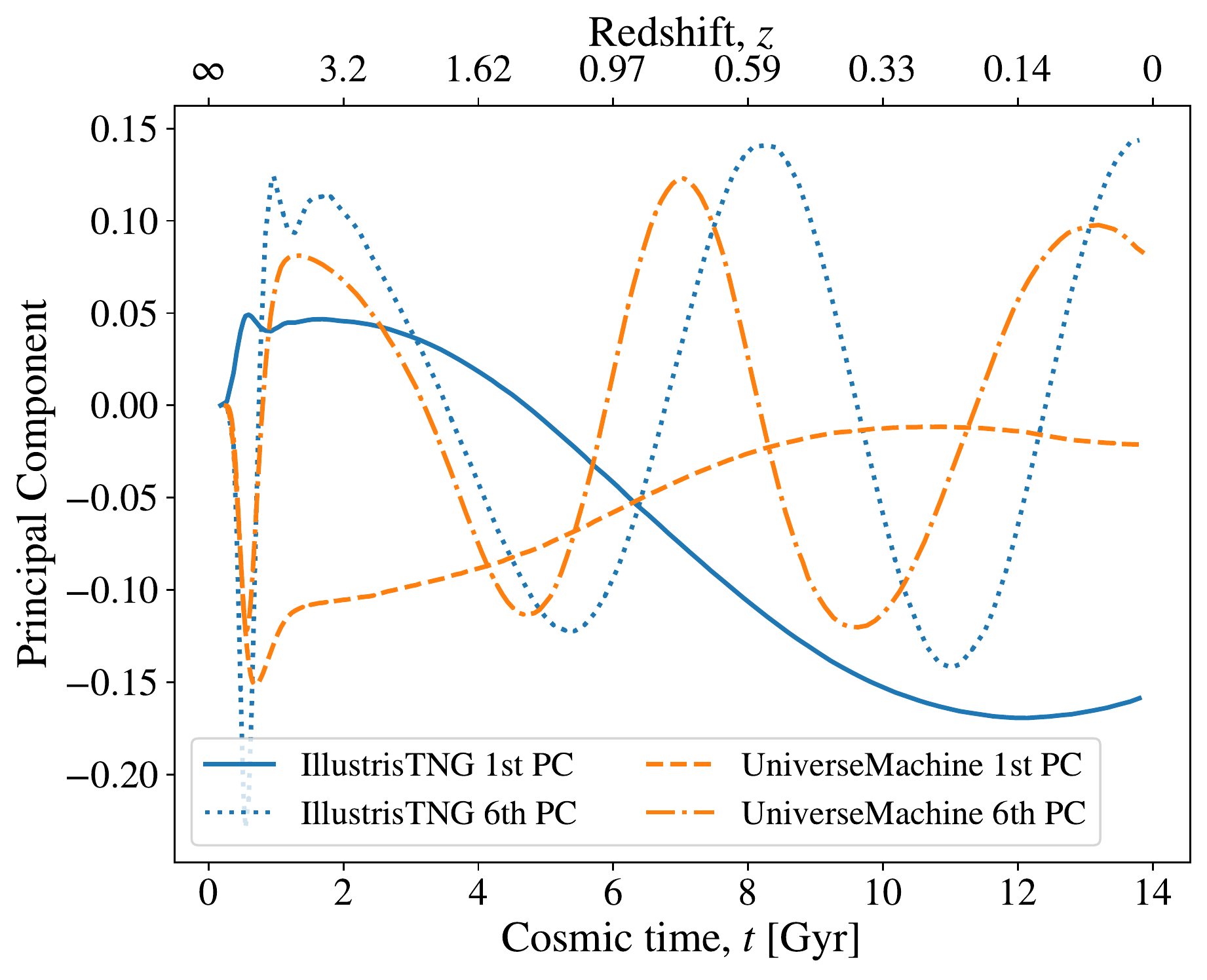}
    \caption{{\bf Left panel.} Performance of PCA-based models in reproducing the SFH of a randomly selected \um galaxy. The blue dotted line indicates the SFH predicted by \um, while the orange and green lines display 5- and 20-PC models, respectively. Although both models capture the overall shape of the \um SFH, only the model with 20 PCs resolves short-term fluctuations. {\bf Right panel.} Principal components resulting from the analysis of \tng and \um SFHs. The solid and dotted (dashed and dot-dashed) lines indicate the first and sixth PC for \tng (\um), respectively. Models using more PCs resolve shorter-term fluctuations because higher-order PCs display an increasing number of oscillations.}
    \label{fig:model_cartoon}
\end{figure*}

\subsection{IllustrisTNG}

\tng is a suite of cosmological hydrodynamical simulations that models the joint evolution of dark matter, gas, stars, and supermassive black holes by incorporating a comprehensive galaxy formation model with radiative gas cooling, star formation, galactic winds, and AGN feedback \citep{Weinberger2017, Pillepich2018a}. Throughout this work, we use publicly available data from the largest hydrodynamical simulation of the suite, TNG300-1 \citep{Nelson2018a}, which evolved $2\,500^3$ gas tracers together with the same number of dark matter particles in a periodic box of $302.6$ Mpc on a side using the moving-mesh code {\sc Arepo} \citep{Springel2010} under {\it Planck} 2015 cosmology. The corresponding mass resolution is 5.9 and $1.1\times10^7\Msun$ for dark matter and gas, respectively, and publicly available galaxy properties are tabulated at 100 snapshots.


\subsection{UniverseMachine}

\um is an empirical model of galaxy formation that maps halo assembly histories onto star formation histories using a set of mass- and redshift-dependent scaling relations \citep[for further details, we refer the reader to][]{behroozi_etal19}. The motivation of considering this model relies on the success of \um reproducing a broad range of galaxy summary statistics across an extensive redshift range, including stellar mass functions, UV luminosity functions, quenched fractions, and two-point clustering.

To generate the \um data that we use in the present work, we run the publicly available code\footnote{\url{https://bitbucket.org/pbehroozi/universemachine/}} on merger trees identified in the Bolshoi Planck simulation with Rockstar and ConsistentTrees \citep{behroozi_etal13a, behroozi_etal13b, klypin_etal16, rodriguez_puebla_etal16}. The Bolshoi-Planck simulation was carried out by evolving $2\,048^3$ dark-matter particles of mass $m_{\rm p}=1.35\times10^{8}M_{\odot}$ on a simulation box of $250\,h^{-1}{\rm Mpc}$ on a side using the ART code \citep{kravtsov_etal97} under cosmological parameters closely matching \citet{planck14b}. After running \um, we end up with $\simeq700\,000$ galaxies with SFHs tabulated at each of the 178 publicly available Bolshoi-Planck snapshots.


\section{SFH models with adjustable levels of variability}
\label{sec:model}

In this section, we use principal component analysis \citep[PCA,][]{Pearson1901} to build approximations to the star formation history of galaxies in \tng and \um. Our PCA-based technique allows us to capture SFH variability over successively shorter timescales by using progressively larger numbers of principal components, which is the key feature we will use to study how physically-motivated levels of SFH variability influences galaxy evolution. We start in \S\ref{sec:model_introduction} by introducing our PCA-based formalism, and in \S\ref{sec:model_pow} we quantify the ability of these models to characterise star formation fluctuations over controllable timescales. Finally, in \S\ref{sec:model_performance}, we study the precision of PCA-based approximations reproducing SFHs predicted by \tng and \um.


\begin{figure*}
    \centering
    \includegraphics[width=\columnwidth]{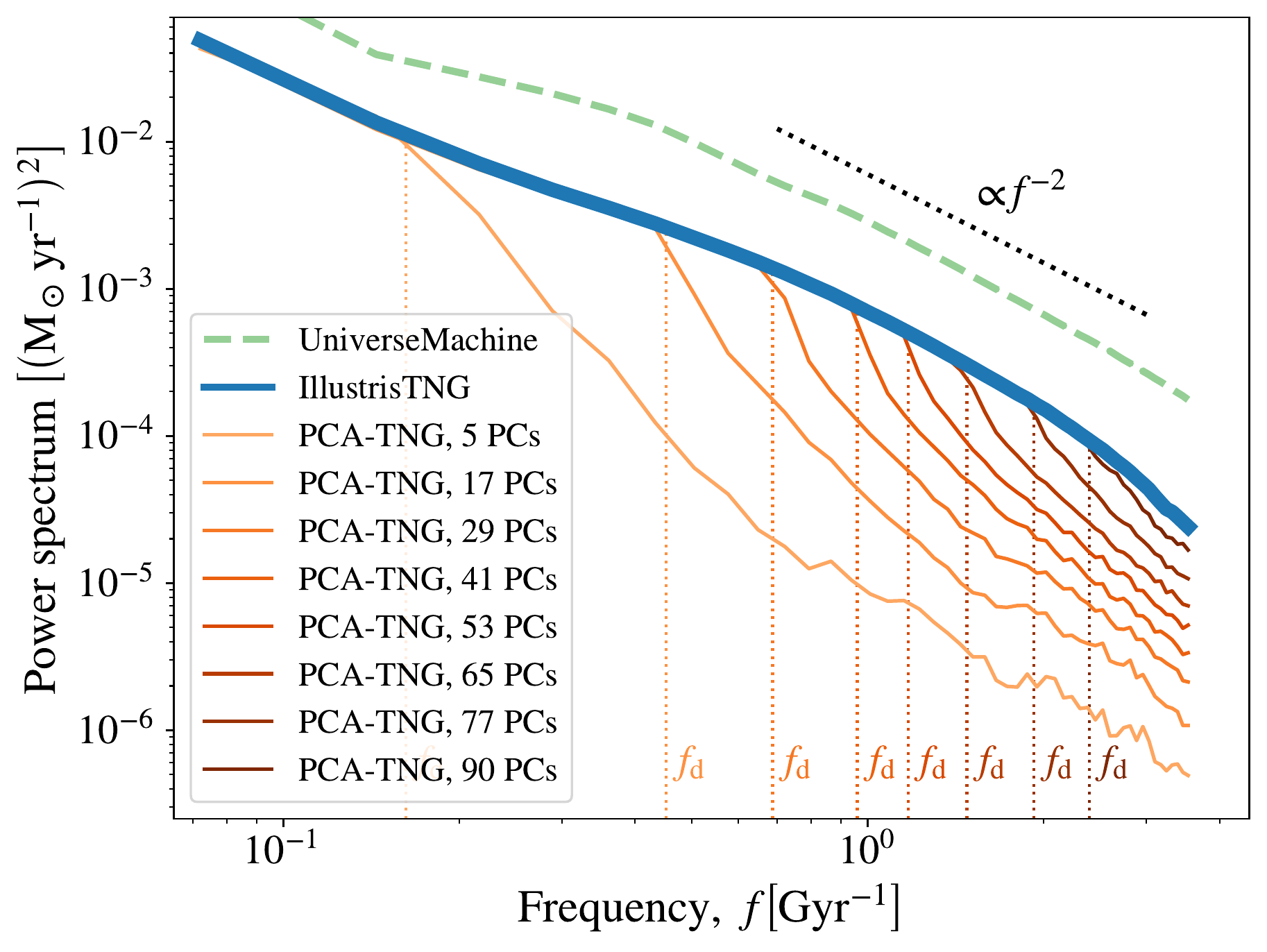}
    \includegraphics[width=\columnwidth]{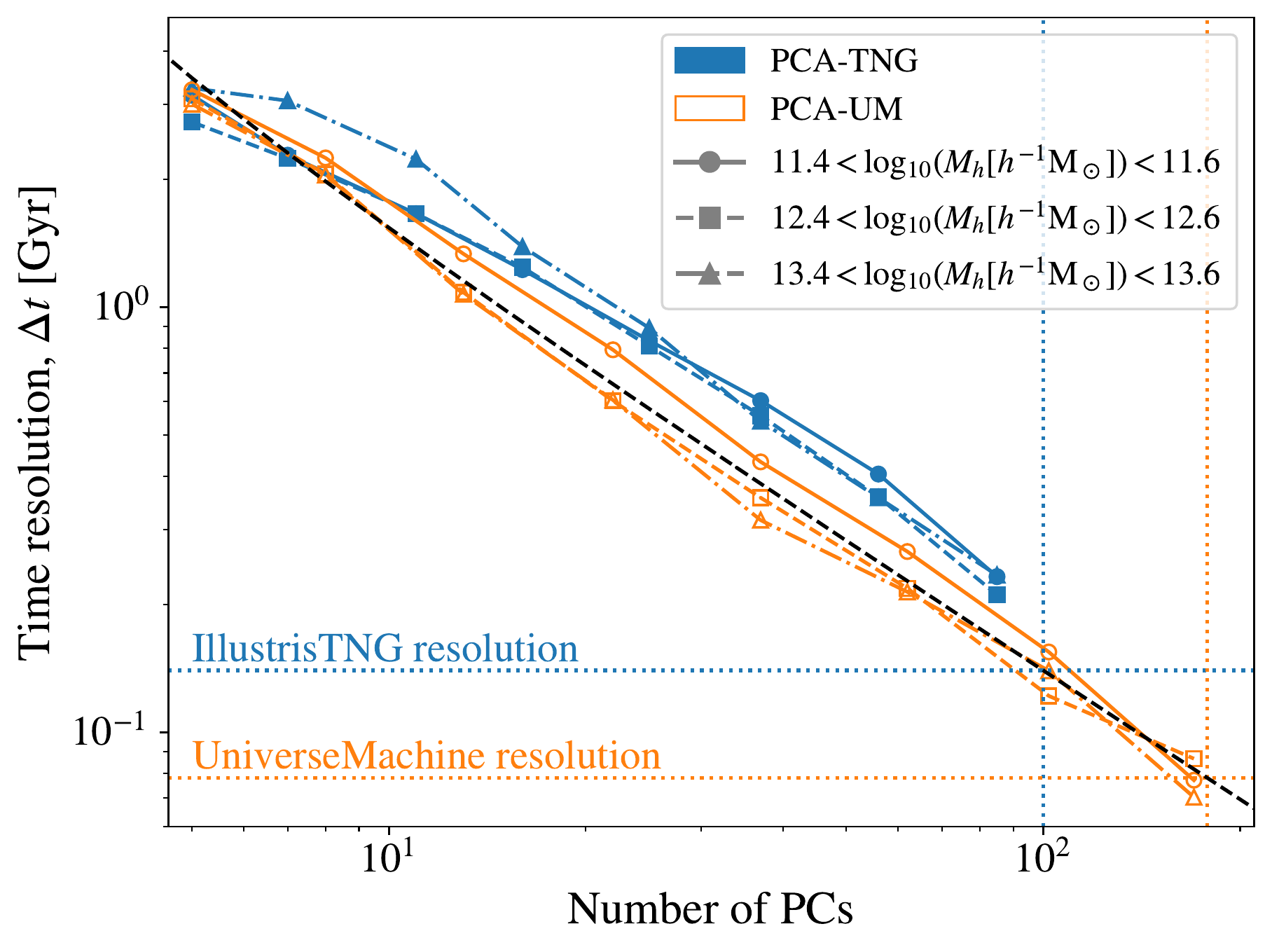}
    \caption{
    {\bf Left panel.} Power spectrum of star formation fluctuations. The blue solid and green dashed lines show median results for \tng and \um, respectively; orange lines display results for PCA approximations to \tng using different numbers of PCs; vertical dotted lines indicate the frequency of the shortest-term fluctuations that each approximation resolves, $f_\mathrm{d}$. {\bf Right panel.} Timescale of the shortest-term fluctuations that PCA-based models resolve, $\Delta t \equiv (2 f_\mathrm{d})^{-1}$. Blue and orange lines display results for \tng and \um, respectively; symbols show results for galaxies hosted by haloes of different mass; the black dashed line depicts the resolution of SFHs tabulated at the same number of cosmic times as the number of PCs indicated at the corresponding point on the x-axis; horizontal dotted lines show the resolution limit of \tng and \um due to the finite number of snapshots in the catalogues. As we can see, PCA-based models using $m$ PCs resolve the spectrum of fluctuations up to the frequency as SFHs tabulated at $m$ cosmic times.}
    \label{fig:model_power_spec}
\end{figure*}

\subsection{Introduction to PCA-based models}
\label{sec:model_introduction}

Principal component analysis is a classical technique in unsupervised machine learning that is widely used in problems of dimensional reduction. In PCA, a multivariate dataset is projected into a coordinate system of orthogonal basis vectors that are referred to as principal components (PCs). This method proceeds by applying successive orthogonal linear transformation in such manner that the first PC accounts for the maximum possible amount of data variance, the second PC maximises the remaining variance under the constraint that it has to be orthogonal to the previous PC, and so on. Since each new PC capture progressively less variance, this technique is ideal for producing SFH models that capture adjustable levels of variability in a dataset. PCA-based models also present additional advantages: they capture the maximum possible amount of variability per unit of information for a linear model, and can reproduce arbitrarily complex SFHs. In this section, we describe our approach to adjust PCA-based models to SFHs predicted by galaxy formation models.

To fit SFHs predicted by \tng and \um, we begin by carrying out a principal component analysis of the SFHs predicted by each of these models separately; this is motivated by the conceivable dependence of PCs on the particularities of galaxy formation models. Specifically, we conduct a PCA of the SFHs predicted by IllustrisTNG-300 and \um evaluated in the Bolshoi-Planck simulation for the 288\,444 and 507\,673 galaxies with $\mspeak>10^9\,\Msun$ and $\mspeak>10^{9.2}\,\Msun$, respectively, where $\mspeak$ is the maximum stellar mass reached by a galaxy over the main branch of its merger tree. In this manner, we end up with $n=100$ and 178 PCs for \tng and \um, respectively, where $n$ is the number of cosmic times at which publicly released IllustrisTNG-300 and Bolshoi-Planck data are tabulated. Finally, we project SFHs onto a number of PCs smaller than $n$, and then we project the resulting linear combination of PCs back into SFHs. As a result, this process results in a PCA-based approximation to each SFH that captures the maximum amount of data variance given a linear combination of $n$ PCs. Note that PCA-based models using more PCs capture an increasingly larger amount of variability, and that a model using $n$ PCs recovers original SFHs without loss of information.

To improve the precision of PCA-based models at fixed number of PCs, we find useful to compute the logarithm of SFHs before carrying out the principal component analysis; this is motivated by the large dynamic range of SFRs predicted by galaxy formation models. To avoid zero-value problems with the logarithm, before performing this operation we assign a minimum SFR of $\dot{M}_*^\mathrm{floor} = 10^{-4}\,\Msunyr$ to galaxies forming stars at any lower rate. We check that the impact of using this floor value is negligible for the \tng and \um galaxies that we consider.

To illustrate PCA-based models, in the left panel of Fig.~\ref{fig:model_cartoon} we display the SFH of a randomly selected \um galaxy, and we show the PCA approximation to this SFH using 5 and 20 PCs. The blue dotted line indicates the SFH predicted by \um, while the orange solid and green dashed lines depict the 5 and 20 PCs models, respectively. As we can see, both models describe the overall shape of the true SFH; however, only the model using 20 PCs resolves short-timescale star formation fluctuations. In the right panel of Fig.~\ref{fig:model_cartoon}, we use solid and dotted (dashed and dot-dashed) lines to display the 1st and 6th PCs resulting from the analysis of \tng (\um) SFHs, respectively; that is, these are two examples of the eigenbasis onto which we projected the SFH shown in the left panel. We find that higher-order PCs display an increasingly larger number of oscillations, which helps visualise why PCA-based models with more PCs resolve shorter variability timescales.


\subsection{Timescale of fluctuations resolved by PCA-based models}
\label{sec:model_pow}

The power spectrum of a time series captures the distribution of variance among the frequency components comprising the signal; consequently, the power spectrum of an SFH describes the amount of variability contained in fluctuations of different frequencies. In this section, we use the power spectra of SFHs to study the performance of PCA-based models in resolving star formation fluctuations of different timescales.

To compute the power spectrum of an SFH, we start by interpolating such SFH into 500 evenly-spaced cosmic times between $t=0$ Gyr and present time. Note that this is a standard approach to compute the power spectrum of unevenly spaced data; we check that the results remain unchanged when using more interpolation points. We continue by computing the fast Fourier Transform of the evenly-spaced SFH, and finally we obtain the power spectrum by taking the square of the absolute value of the transform.

In the left panel of Fig.~\ref{fig:model_power_spec}, we display the median power spectrum of \tng and \um SFHs as well as the power spectrum of different PCA-based approximations to \tng SFHs. The blue solid and green dashed lines depict the results for \tng and \um, respectively, while orange lines do so for PCA-based models to \tng SFHs using a different number of PCs. Vertical dotted lines indicate the frequency at which the relative difference between the power spectrum of \tng SFHs and of PCA-based approximations reaches $20\%$; in what follows, we refer to this quantity as damping frequency, $f_\mathrm{d}$. As we can see, the power spectrum of PCA-based models agrees with that of \tng galaxies for frequencies lower than $f_\mathrm{d},$ and models using more PCs present increasingly higher $f_\mathrm{d}$. Taken together, these results let us conclude that PCA-based models using more PCs reproduce the full spectrum of star formation fluctuations up to higher frequencies. We have verified that these findings hold for a range of other definitions of $f_\mathrm{d}$.

We can also see that the power spectrum of \tng and \um SFHs decreases with frequency as a power law; to help visualise this trend, in the left panel of Fig.~\ref{fig:model_power_spec} we use a dotted line to indicate a power spectrum inversely proportional to the square of the frequency. The main consequence of this dependency is that the integrated amount of variability on long-term fluctuations is much larger than that on short-term fluctuations. As a corollary, the level of variability encoded in SFH fluctuations not resolved by the galaxy catalogue due to the limited number of simulation snapshots must be subdominant relative to the variability arising from well-resolved fluctuations.

In principle, the frequency dependence of SFH power spectra could depend on the treatment of baryonic processes by galaxy formation models. Notably, however, different state-of-the-art models predict quite similar frequency dependence \citep{Iyer2020}, including five different cosmological hydrodynamical simulations, three suites of zoom simulations, a semi-analytic model, and an empirical model. Consequently, our findings likely pertain, at least qualitatively, to other galaxy formation models beyond \tng and \um. Interestingly, the power spectrum of observational data also shows similar frequency dependence \citep{Wang_Lilly_2020}, indicating that long-term fluctuations also dominate star formation variability for observed galaxies.

On the other hand, we find that the power spectrum of \um SFHs presents a larger overall {\em amplitude} relative to that of \tng SFHs, and thus star formation fluctuations predicted by \um show larger amplitudes than those predicted by \tng. As a result, the added contribution of all fluctuations not resolved by a PCA-based model with $m$ PCs is larger for \um than for \tng. We return to this point in \S\ref{sec:model_pow} when we present a metric quantifying the burstiness of a model in terms of the cumulative contribution of the model's unresolved SFH fluctuations.

In the right panel of Fig.~\ref{fig:model_power_spec}, we show the timescale of the shortest fluctuations that PCA-based models resolve precisely, $\Delta t \equiv (2 f_\mathrm{d})^{-1}$. Blue and orange lines display the results for PCA-based approximations to \tng and \um SFHs, respectively; symbols show results for galaxies hosted by haloes of different mass, and the black dashed line depicts the resolution of SFHs tabulated at the same number of cosmic times as PCs in the x-axis. We show the results as a function of host halo mass because \tng and \um galaxies in more massive haloes reach the peak of their SFH faster, present a larger stellar mass, and quench at earlier times (CH20); as a result, the performance of PCA-based models resolving variability could depend upon host halo mass. Nonetheless, we find that the performance of these models is quite comparable for different host halo masses at a fixed number of PCs. In addition, we can readily see that the results for \tng and \um are compatible, providing an alternative demonstration of the similar dependence of variability on frequency for physically-motivated SFHs.

In the right panel of Fig.~\ref{fig:model_power_spec}, we can also see that PCA-based models using $m$ PCs resolve the spectrum of fluctuations up to a similar frequency as SFHs tabulated at $m$ cosmic times; this is because the number of oscillations of a PC increases with its order (see \S\ref{sec:model_introduction}). This result is significant because it establishes a rough equivalence between models with $m$ PCs and piecewise models using SFHs tabulated at $m$ cosmic times, the latter being in comparatively more common use in the literature.


\begin{figure}
    \centering
    \includegraphics[width=\columnwidth]{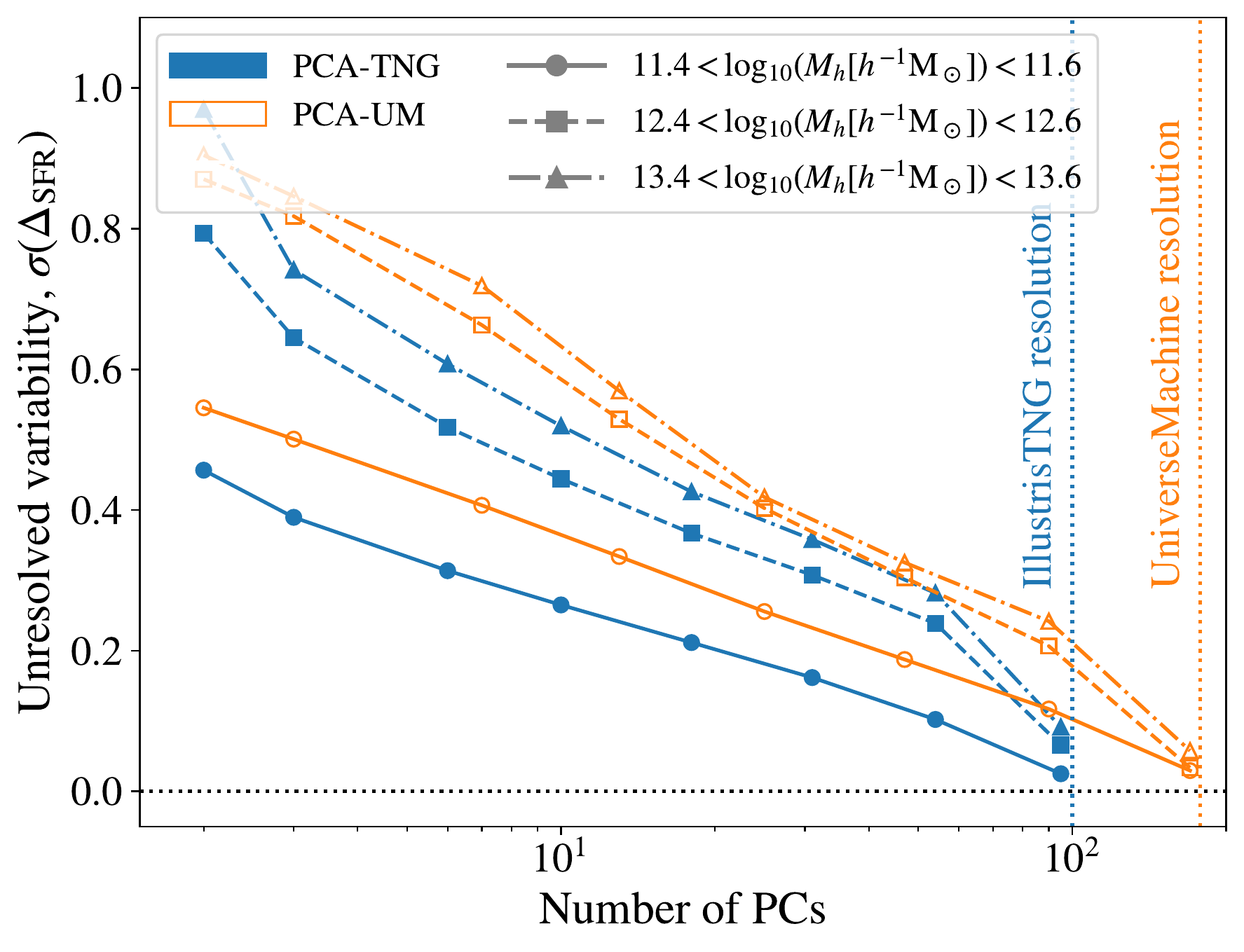}
    \caption{Level of unresolved SFH variability for different PCA-based models (see Eq.~\ref{eq:sigmasfr} for the definition of the vertical axis). Blue and orange colours show results for PCA approximations to \tng and \um SFHs, respectively; results for galaxies in haloes of different masses are shown with different points as indicated in the legend. At fixed number of PCs, the level of unresolved variability varies across samples because the amplitude of star formation fluctuations depends on both host halo mass and the particularities of galaxy formation models.}
    \label{fig:pca_expvar}
\end{figure}

\begin{figure}
    \centering
    \includegraphics[width=\columnwidth]{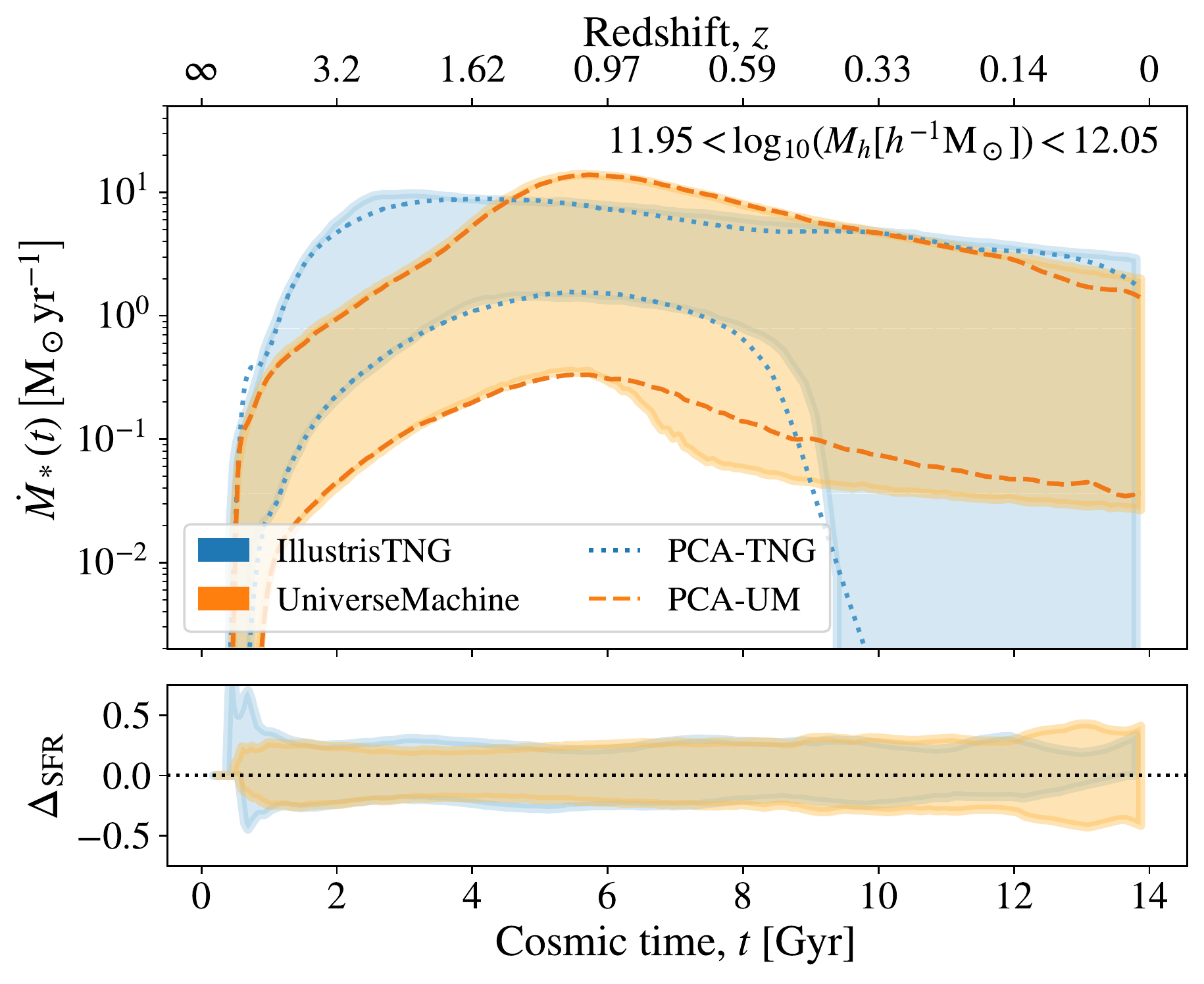}
    \caption{SFHs predicted by \tng and \um for galaxies hosted by $\logMpeak\simeq12$ haloes, displayed in comparison to their PCA-based approximations. Blue and orange shaded areas display the 16-to-84 percentile region for \tng and \um SFHs, respectively, while blue and orange lines show the same quantity for PCA models with 6 and 22 PCs, each of which present $\sigmasfr=0.35$ relative to their associated simulation. The bottom panel shows the logarithmic difference between the exact and PCA-based SFHs. Even though the two SFH models are based on a different number of components, each approximation has the same $\sigmasfr,$ and thereby shows analogous performance in reproducing SFHs, a simple demonstration of the utility of this metric.}
    \label{fig:pca_performance}
\end{figure}

\subsection{Performance of PCA-based models reproducing SFHs}
\label{sec:model_performance}

In the previous section, we showed that PCA-based methods using larger numbers of PCs resolve the full spectrum of star formation fluctuations up to higher frequencies. We confirmed previous results that the frequency dependence of SFH power spectra is very similar amongst contemporary galaxy formation models, but we found that the amplitude of such spectra depends on the particularities of the model. This suggests that when quantifying the overall performance of any SFH approximation, it should be a reasonable approximation to restrict consideration to the timescale of the shortest fluctuations that the approximating model can resolve, and the integrated contribution of all unresolved fluctuations.

Guided by this perspective, in this section we define a metric that we will subsequently use to quantify how accurately an SFH model recovers the true underlying star formation history of a galaxy. To quantify the performance of an SFH approximation, we consider the standard deviation of the logarithmic difference between the original SFHs and PCA-based approximations,
\begin{eqnarray}
\label{eq:sigmasfr}
\sigmasfr\equiv\mathrm{std}[\log_{10}(\dot{M}_*^\mathrm{orig}) - \log_{10}(\dot{M}_*^\mathrm{model})],
\end{eqnarray}
where $\mathrm{std}$ stands for standard deviation. Models showing better precision, i.e. resolving more variability, present smaller values of $\sigmasfr$. Hereafter, when discussing the success of a star formation history model, we will refer to the quantity $\sigmasfr$ as the level of unresolved variability of the model, and we will informally use the term ``burstiness'' to refer to short-term variability in SFH. By design, this summary statistic is independent of the number of cosmic times at which SFHs are tabulated, and thus it can be used to compare the variability of SFHs predicted by different simulations and galaxy formation models.

In Fig.~\ref{fig:pca_expvar}, we display the level of unresolved variability for models using different numbers of PCs. Blue and orange curves show results for PCA approximations to \tng and \um SFHs, respectively; different symbols show results for galaxies in haloes of different mass, as indicated in the legend. All curves in the figure show that $\sigmasfr$ decreases monotonically with the number of PCs: using more principal components enables the approximating model to capture fluctuations up to shorter timescales. As we can see, the performance of PCA-based models varies across samples, reflecting the dependence of the amplitude of star formation fluctuations on the physics of galaxy formation. We can also see that the value of $\sigmasfr$ increases with host halo mass at a fixed number of PCs; this is just a manifestation of the increasing dynamical range of SFR values for galaxies hosted by more massive haloes in \tng and \um (CH20).

Interestingly, at fixed number of components, the PCA approximation shows better precision in reproducing SFHs of \tng relative to \um. We can understand this result in terms of Fig.~\ref{fig:model_power_spec}: the power spectrum of \tng SFHs presents smaller amplitude relative to that of \um SFHs (see \S\ref{sec:model_pow}). Based on these results, our PCA-based analysis allows us to conclude that \um predicts greater levels of burstiness in comparison to \tng. As an alternative manifestation of this observation, PCA-based approximations to \tng and \um SFHs that use 6 and 22 PCs, respectively, present equivalent performance in reproducing SFHs. Thus, the number of degrees of freedom of an SFH model is less indicative of its success relative to the value of $\sigmasfr,$ motivating our choice of $\sigmasfr$ as a standard reference for comparing the performance of parametric and non-parametric models in their ability to reproduce any type of SFH.

Finally, in Fig.~\ref{fig:pca_performance}, we show the variety of SFHs for galaxies hosted by $\logMpeak\simeq12$ haloes. Blue and orange shaded areas display the 16-to-84 percentile region for \tng and \um SFHs, respectively, while blue and orange error bars show PCA-based models presenting $\sigmasfr=0.35$. In the bottom panel, we show the logarithmic difference between the original SFHs and the PCA-based approximations. As we can see, PCA-based models describe individual SFHs in an unbiased fashion, as well as faithfully reproduce the diversity of the target SFHs. Moreover, the precision of these models reproducing \tng and \um SFHs is approximately the same, confirming that $\sigmasfr$ captures the level of unresolved burstiness in the data independently of the shape of the analysed SFHs and the number of cosmic times at which these SFHs are tabulated.


\begin{figure}
    \centering
    \includegraphics[width=\columnwidth]{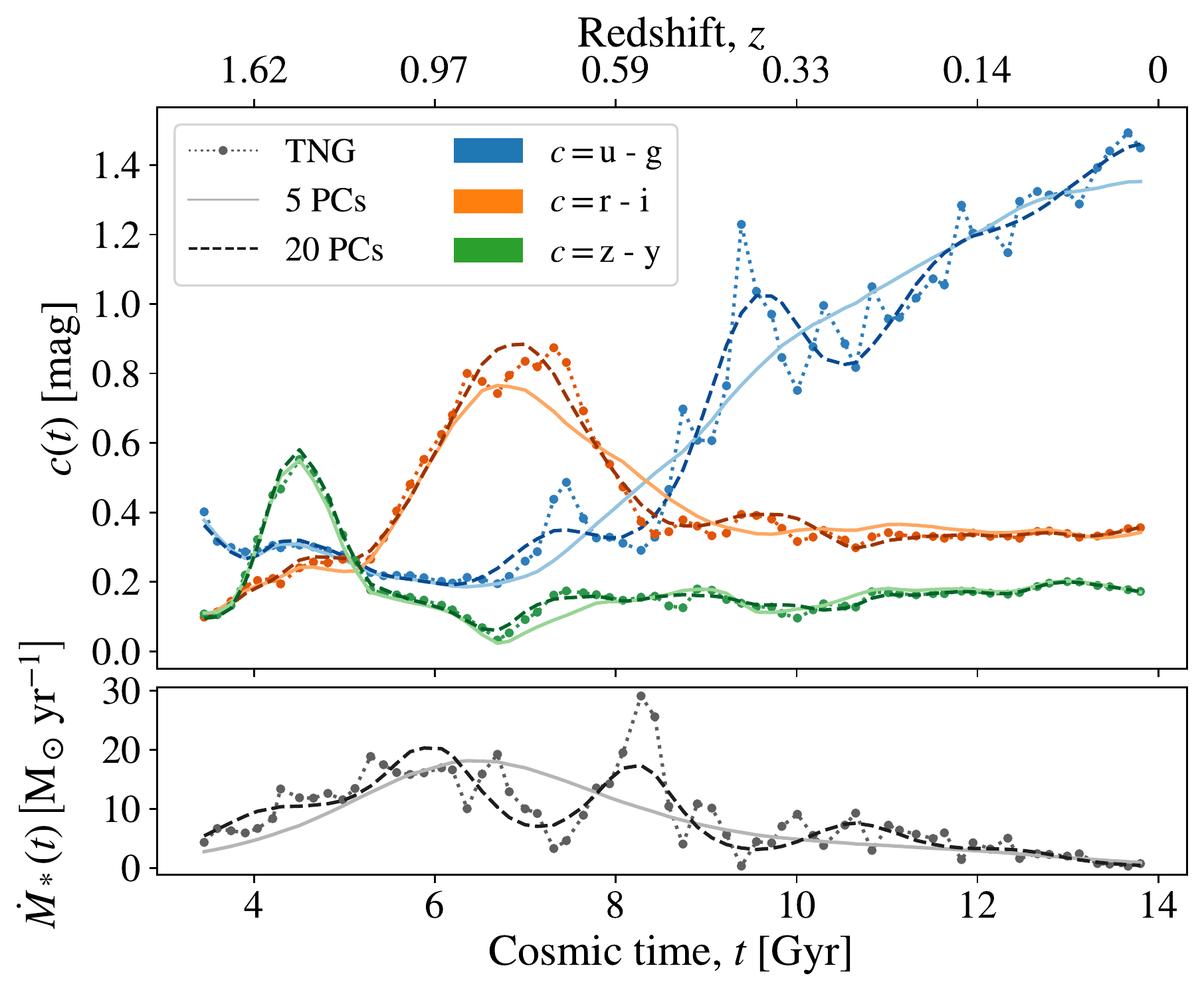}
    \caption{Redshift evolution of the u-g, r-i, and z-y broad-band colours (top panel) and SFR (bottom panel) of a randomly selected \tng galaxy. Dotted lines show the \tng SFH and colours produced using this SFH; solid and dashed lines indicate PCA-based approximations to the \tng SFH using 5 and 20 PCs, respectively, and colours generated using these approximations. This individual example illustrates the general principles that positive and negative star formation fluctuations induce colour variations towards bluer and redder values, respectively, and that SFH approximations with better time resolution reproduce galaxy colours more faithfully.}
    \label{fig:colours_cartoon}
\end{figure}

\section{Influence of star formation variability on galaxy colour}
\label{sec:colours}

In \S\ref{sec:model}, we introduced our PCA-based formalism for approximating the star formation history of galaxies, and we showed how or methodology allows us to quantify burstiness, i.e., short-term variability in SFH. In this section, we use our formalism to study how star formation variability influences observed-frame broad-band colours across time. We begin in \S\ref{sec:colours_intro} with an overview of the relationship between SFH and galaxy colour using a specific simulated galaxy as an illustrative example. Then, in \S\ref{sec:colours_burstiness}, we proceed to quantify how physically-motivated levels of burstiness influence galaxy colours. Finally, in \S\ref{sec:colours_properties}, we compare the relative importance of burstiness to short-term variability in other galaxy properties such as metallicity and extinction from dust.


\subsection{Qualitative analysis}
\label{sec:colours_intro}

We begin our analysis by using the stellar population synthesis code \fsps \citep{Conroy2009, conroy_gunn_2020}, together with its python interface \citep{python_fsps}, to produce broad-band colours through LSST-like filters using both the SFHs directly predicted by \tng and \um, and PCA-based approximations to these SFHs. Following the same approach as in CH20, we generate colours while holding fixed all parameters of the SPS model except the input SFH, allowing us to isolate the impact of SFH on colours from that of other galaxy properties. We then study the impact of burstiness on colours by comparing galaxy colours produced using SFHs with different levels of variability. For the fixed \fsps parameters, we use the following configuration: Chabrier initial mass function \citep{Chabrier2003}, stellar and gas metallicities equivalent to solar values, dust attenuation of light coming from stars younger and older than 10 Myr equal to $\tau_{V}^\mathrm{BC}=1$ and $\tau_{V}^\mathrm{ISM}=0.3$ \citep{Charlot2000}, respectively, power-law dust attenuation curve of index $n_\tau=-0.7$ \citep{Charlot2000}, and the zero-age ionisation at the Str\"omgren radius set to $\log_{10} U_S=-2$.

Following this strategy, we produce colours at every redshift at which SFHs are tabulated in the publicly available catalogues; throughout the remainder of this work, we refer to the resulting time series of galaxy colours as ``colour histories.'' In the top panel of Fig.~\ref{fig:colours_cartoon}, dotted lines display the u-g, r-i, and z-y colour histories produced using the SFH of a randomly selected \tng galaxy, while solid and dashed lines show the corresponding colour histories produced using PCA-based approximations with 5 and 20 PCs, respectively. Using black curves with same line-style conventions, in the bottom panel of this figure we display the \tng SFH and its corresponding PCA-based approximations. As we can see, the reproduction of the broad characteristics of this example galaxy's colour history is reasonably accurate even for the 5-PC model that only very coarsely captures the underlying SFH. We explore the connection between unresolved variability and colour precision more quantitatively in \S\ref{sec:colours_burstiness} below.

As we can see, the departure of the z-y, r-i, and u-g colour histories toward redder values at cosmic times $t=4,$ 5.5, and 8.5 Gyr is not a direct manifestation of fluctuations in the colour history. This effect is just a direct consequence of studying observed-frame colours: these features reflect the transition of the $4000\,\angstrom$ break across LSST filters. The $4000\,\angstrom$ break is caused by the rapid onset of stellar photospheric opacity shortward of $4000\,\angstrom$, and it appears as a step in the spectral energy distribution of a galaxy \citep{ohman1934_SpectrographicStudiesRed, hamilton1985_SpectralEvolutiongalaxies}, presenting an increasing size for galaxies with later spectral type. At $z=0$, the left and right sides of this feature are inside the u and g filters, respectively, resulting in a red u-g colour. As redshift increases, the left side of the $4000\,\angstrom$ break progressively enters the g filter, making the u-g colour increasingly bluer. This feature ultimately leaves the g filter at $z\simeq0.5$; consequently, the u-g colour does not evolve strongly at even higher redshifts. The same explanation applies to the bumps shown by the r-i and z-y colour histories.

Independently of the evolution of colour histories due to the redshift of spectral features, we can see that positive and negative star formation fluctuations apparently trigger colour variations towards bluer and redder values, respectively. We study this in detail in Appendix~\ref{app:response}, finding that star formation fluctuations cause a response on all LSST colours faster than the minimum timescale resolved by \tng and \um SFHs, which is approximately $0.14$ Gyr. Furthermore, we find that for such timescales, positive and negative star formation fluctuations induce colour fluctuations towards bluer and redder values, as expected, and that the impact of these fluctuations on colours is more substantial for increasingly shorter wavelengths.

On basic physical grounds, we of course expect star formation fluctuations to induce variations in galaxy colours because the number of O and B stars increases after a star-forming episode, and these stellar types dominate galactic light at short wavelengths \citep{Renzini1986, Charlot1991}. Thus, it is natural that an increase and decrease in the number of O and B stars results in galaxy colours turning bluer and redder, respectively, and that colours with shorter wavelengths are increasingly more affected. Additionally, these stellar types dominate blue galactic light for only a few hundred million years due to the rapid departure of O and B stars from the main sequence, explaining the synchronicity between fluctuations in star formation and colour.



\begin{figure}
    \centering
    \includegraphics[width=\columnwidth]{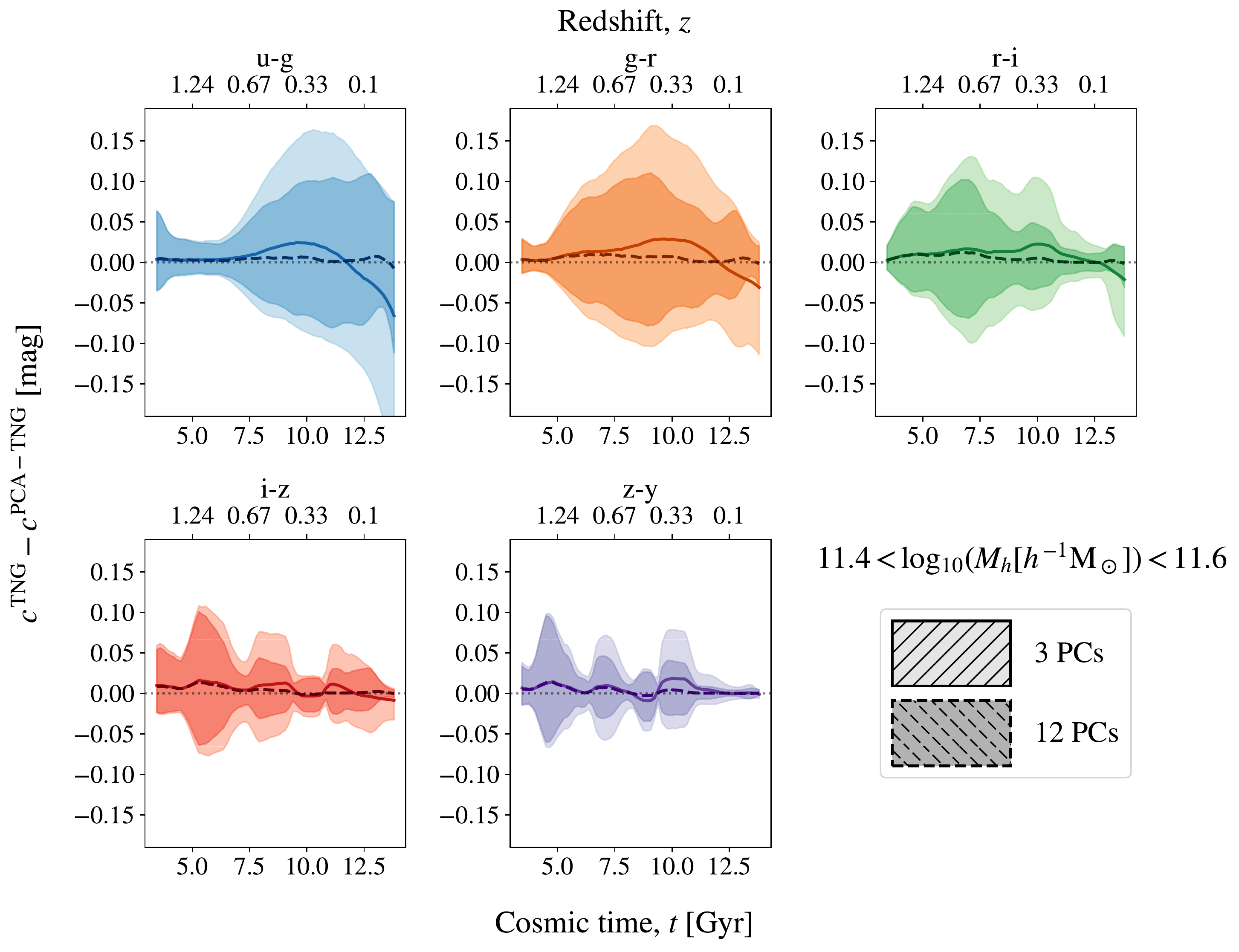}
    \caption{
    Difference between colours produced using exact and PCA-approximated \tng SFHs. Solid lines and light-shaded areas (dashed lines and dark-shaded areas) indicate the median and the 16-to-84 percentile region of the difference between colours produced using exact SFHs and 3-PC (12-PC) models, respectively; each panel shows the results for the broad-band colour indicated at the top of the panel. As expected, the agreement between colours produced using \tng SFHs and 12 PC approximations is better than that between colours generated using \tng SFHs and 3 PC approximations, although even in the 3 PC case, we note that imprecision is limited to the level of $\sim0.15$ mag.}
    \label{fig:colours_diff_burstiness}
\end{figure}

\begin{figure}
    \centering
    \includegraphics[width=\columnwidth]{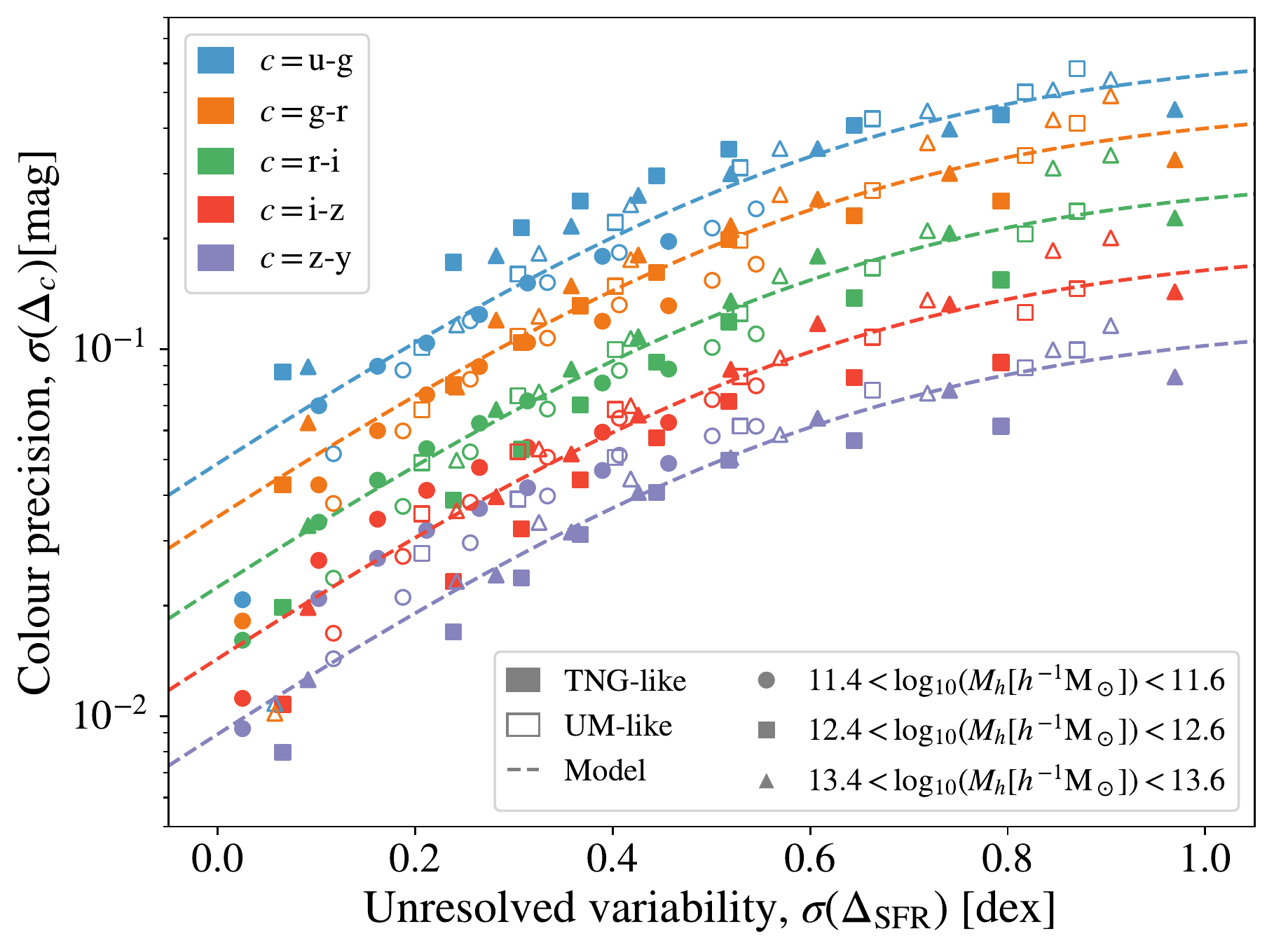}
    \caption{
    Precision of broad-band colours produced using SFH models resolving different levels of variability. Filled (open) symbols indicate the standard deviation of the difference between colours produced using \tng (\um) SFHs and PCA-approximations, respectively. Circles, squares, and triangles show results for galaxies hosted by $\logMpeak=11.5,$ 12.5, and 13.5 haloes; different colours display the results for distinct broad-band colours as indicated in the legend. Each dashed line shows the fitting function presented in Eq.~\ref{eq:evar_col}, which provides a scaling relation between the typical level of error in the SFH approximation and the associated imprecision in predictions for colour. The precision of colours generally increases for SFH models resolving more variability; as we can see, our fitting function faithfully captures this trend. Therefore, this function should be useful in assessing the accuracy required by an SFH model to achieve some target precision in predictions for broad-band colours.}
    \label{fig:colours_precision}
\end{figure}

\subsection{Quantitative analysis}
\label{sec:colours_burstiness}

In the last section, we showed that positive and negative star formation fluctuations induce variations in broad-band colours towards bluer and redder values, respectively. Therefore, we expect SFH models capturing more variability to result in more precise galaxy colours. More quantitatively, in this section we study the colours produced directly with \tng and \um SFHs, and we compare these colours to those produced using PCA-based approximations that resolve different levels of burstiness.

In Fig.~\ref{fig:colours_diff_burstiness}, we focus on the colours of galaxies in \tng whose underlying halo mass lies in the range $11.4<\logMpeak<11.6.$ For every galaxy in this sample, we calculate the colour history produced using the exact \tng SFH, as well as the colour history resulting from the approximate SFH based on PCA. We plot the difference between the two on the vertical axis, showing results for different colours according to the title of each panel. Solid lines show the median difference; light-shaded (dark-shaded) areas show the 16-to-84 percentile region of this colour difference for PCA-based models with 3 (12) components. As we can, the approximate model with 12 PCs reproduces the exact colours without appreciable bias, but the colours produced using the 3-PC model show some level of redshift-dependent offset.

One of the most prominent features of the figure is that the colour precision depends on both wavelength and redshift at a fixed number of PCs. This feature reflects both the sensitivity of galaxy light to the fluctuating abundance of O and B stars and the transition of the $4000\,\angstrom$ break across filters. The strength of the $4000\,\angstrom$ break increases rapidly as redder stellar types dominate the light of a galaxy; therefore, when the $4000\,\angstrom$ falls in one of the two bands defining a colour, this colour is particularly susceptible to star formation fluctuations.

In Fig.~\ref{fig:colours_precision}, we quantify how unresolved SFH variability translates into imprecision in predictions for galaxy colour. We quantify unresolved SFH variability on the horizontal axis using $\sigmasfr,$ the standard deviation of the logarithmic difference between exact and approximate star formation histories, the same metric developed in \S\ref{sec:model_performance} to evaluate the success of an SFH approximation (see also Fig.~\ref{fig:pca_expvar}). On the vertical axis we quantify imprecision in colour using $\sigma(\Delta_c),$ the standard deviation of the difference between approximate and exact colour histories. Filled and open symbols indicate the results for \tng and \um SFHs, respectively, circles, squares, and triangles show results for galaxies hosted by $\logMpeak=11.5$, 12.5, and 13.5 haloes, and different coloured points display the results for distinct broad-band filters. As expected by the physical picture outlined above, at fixed $\sigmasfr$ the precision of galaxy colours improves for longer wavelengths. We also find that the average difference between exact and PCA-approximated colours is compatible with zero within error bars; therefore, colours produced using PCA-based models do not show a significant bias in colour relative to those generated using the exact SFHs from these simulations.

We can readily see in Fig.~\ref{fig:colours_precision} that the precision of different PCA-based models presents minimal dependency on either host halo mass or the particularities of galaxy formation models at a fixed unresolved variability. Given that SFHs predicted by \tng and \um present significant differences as well as strong dependence on host halo mass (CH20), this result is providing further evidence that $\sigmasfr$ captures the level of unresolved variability in the data precisely. Furthermore, it motivates exploration of a scaling relation between colour precision and $\sigmasfr$. The trends shown in the figure suggest considering the following functional form,
\begin{eqnarray}
    \label{eq:evar_col}
    \sigmacol = \frac{a_0 + a_1\lambda_\mathrm{cen} + a_2\lambda_\mathrm{cen}^2}{1+\exp\{-k[\sigmasfr-\sigma_0]\}},
\end{eqnarray}
where $\lambda_\mathrm{cen}$ refers to the average central wavelength of the two bands that define each broad-band colour, while $a_0=1.57\,\mathrm{mag}$, $a_1=-2.67\,\mathrm{mag}\,\umu\mathrm{m}^{-1}$, $a_2=1.19\,\mathrm{mag}\,\umu\mathrm{m}^{-2}$, $k=4.27$, and $\sigma_0=0.59$ are the best-fitting parameters of the model to data.

In Fig.~\ref{fig:colours_precision}, we plot this fitting function using dashed lines. As we can see, Eq.~\ref{eq:evar_col} captures the dependency of colour precision on unresolved burstiness quite faithfully, and so we will use this scaling relation to estimate the precision of galaxy colours produced using any SFH model (see also \S\ref{sec:forward}). We emphasise that our fitting function only provides the {\em average} precision with which some approximate SFH model is able to {\em statistically} capture colour history, and that colour precision of any approximating model will necessarily exhibit redshift-dependent differences from the truth when evaluated for any particular galaxy.

\begin{figure}
    \centering
    \includegraphics[width=\columnwidth]{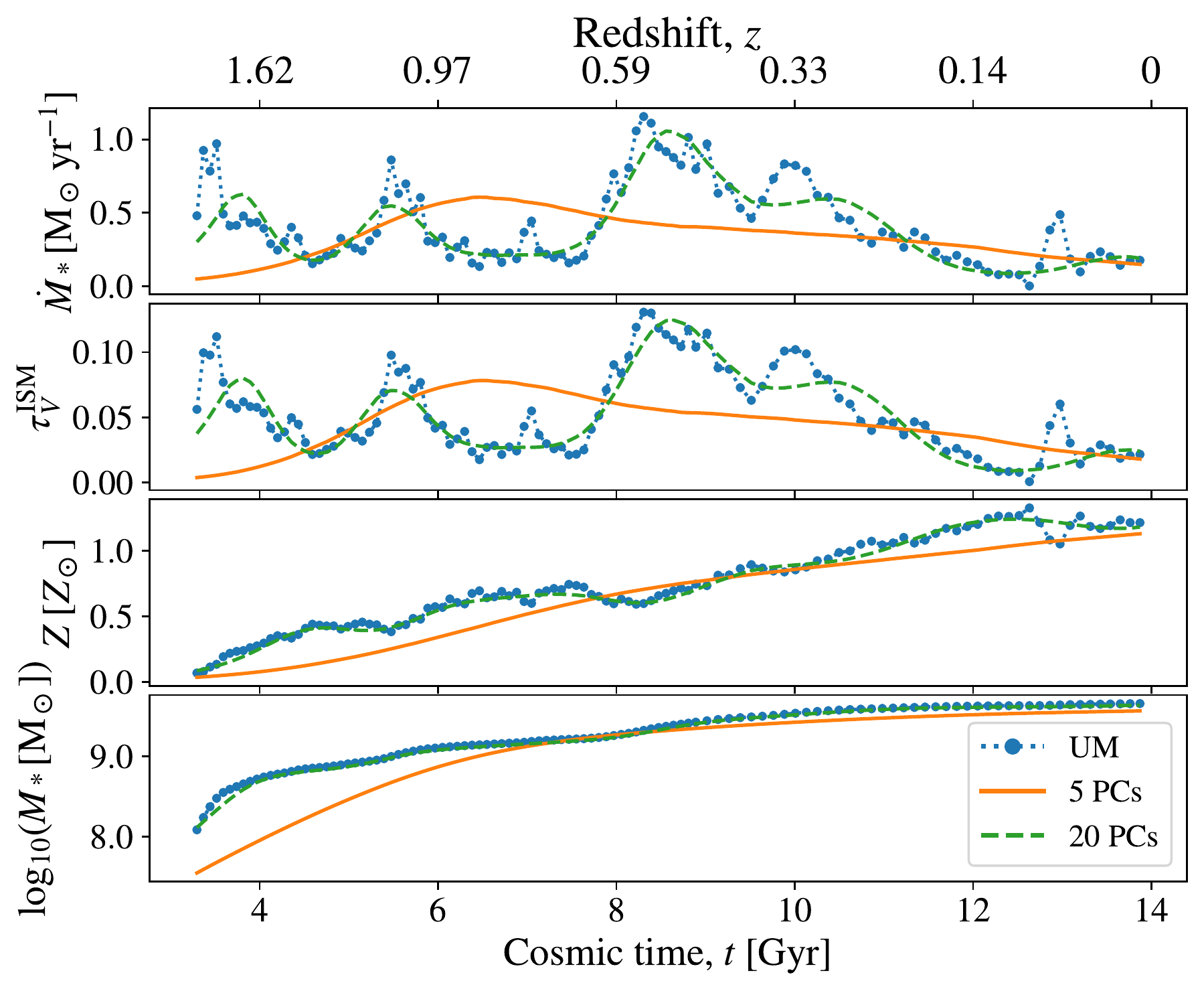}
    \caption{Redshift evolution of SFR (top), dust attenuation (top middle), metallicity (bottom middle), and stellar mass (bottom) for a randomly selected \um galaxy. Blue lines indicate the results for the SFH taken directly from \um, while orange and green lines show results for the PCA-based approximations based on 5 and 20 PCs, respectively.}
    \label{fig:colours_gal_properties}
\end{figure}

\begin{figure}
    \centering
    \includegraphics[width=\columnwidth]{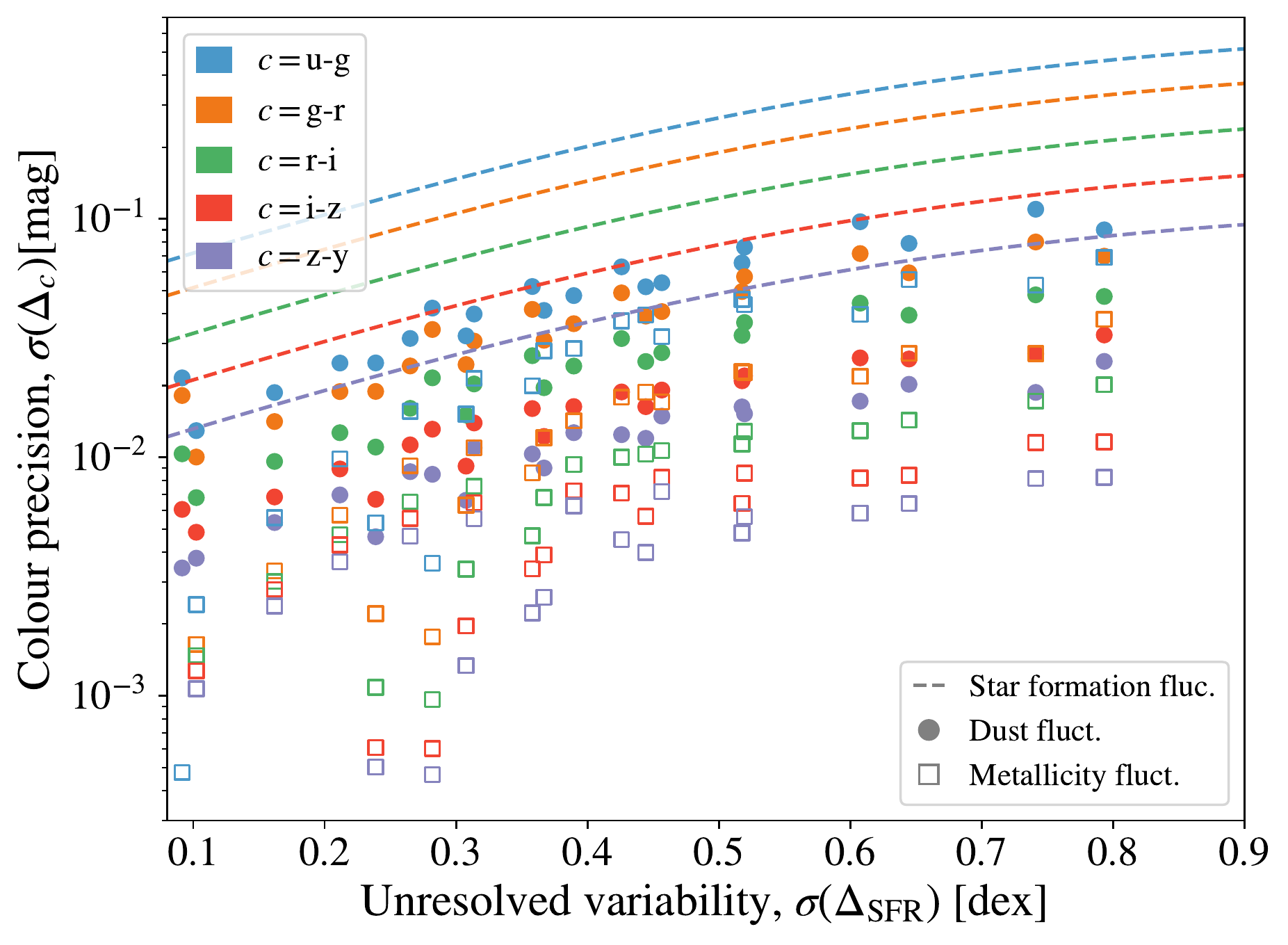}
    \caption{Impact of metallicity and dust attenuation variability on galaxy colours. The vertical axis shows the imprecision in predictions for galaxy colour associated with the level of unresolved variability shown on the horizontal axis. Circles and squares indicate the influence of unresolved short-term fluctuations in dust attenuation and metallicity, respectively, and dashed lines show the impact of unresolved star formation variability. By comparing circles and squares to the dashed line of the same colour, we can see that the influence of star formation variability on all colours is much stronger relative to the impact of fluctuations in dust attenuation and metallicity.}
    \label{fig:colours_precision_properties}
\end{figure}

\subsection{Impact of fluctuations in dust and metallicity on colours}
\label{sec:colours_properties}

As noted at the beginning of \S\ref{sec:colours}, for all our previous results we have exclusively considered how star formation variability influences galaxy colours. In particular, for all calculations in \S\ref{sec:colours_intro} and \S\ref{sec:colours_burstiness}, both stellar metallicity and dust attenuation were held fixed to the constant values assumed by our fiducial model described at the beginning of \S\ref{sec:colours_intro}. Nonetheless, basic physical considerations lead to the natural expectation that star formation variability induces fluctuations in both metallicity and dust extinction, and so in this section we study how such concomitant variations manifest in the observed colours of galaxies.

The metallicity of a galaxy depends upon the complex interplay between metal enrichment, gas inflows, and gas outflows, with freshly infalling gas typically presenting low metallicity, and outflows commonly exhibiting characteristics of metal-rich gas \citep[e.g.,][]{finlator2008_OriginGalaxymass, dave2011_GalaxyEvolutioncosmological, lilly2013_GASREGULATIONGALAXIES, dave2017_MufasaGalaxystar, Torrey2019}. Since the infall of pristine gas should induce star formation activity, and the explosion of supernovae should produce outflows, it is physically natural to expect correlations between star formation and metallicity, and by now a wide variety of observations indicate that this expectation is borne out \citep[e.g.,][]{ellison2008_CLUESORIGINMASSMETALLICITY, lara-lopez2010_FundamentalPlanefield, Mannucci2010}. At fixed stellar mass, galaxies with higher SFR present increasingly lower gas metallicity, a trend \citet{Mannucci2010} found to be well-captured by the following relation:
\begin{eqnarray}
    \label{eq:met_model}
    12 + \log_{10}(\mathrm{O}/\mathrm{H}) =&\,8.90 + 0.37\,m - 0.14\,s - 0.19\,m^2 \nonumber\\
    &\hspace*{-1.3cm}+\ 0.12\,m\,s - 0.054\,s^2,
\end{eqnarray}
where $m=\log_{10}(M_*[10^{10}\Msun])$ and $s=\log_{10}(\dot{M}_*[\Msunyr])$. This expression was calibrated in \citet{Mannucci2010} using an observational sample of star-forming galaxies; to avoid extrapolation problems for quenched galaxies, when we apply this model to galaxies with SFR smaller than $s_\mathrm{min}=1.13\,m + 1.28,$ we set the metallicity of these to $\log_{10}(\mathrm{O}/\mathrm{H})(m, s_\mathrm{min})$. The previous relation captures the strong correlation between gas-phase metallicity, stellar mass, and star formation rate, predicting concurrent gas metallicity and star formation fluctuations.

The SED of a galaxy depends primarily on the metallicity of its stars \citep[e.g.,][]{Conroy2013}, with changes in gas-phase metallicity mainly affecting the properties of galaxy emission lines. Hydrodynamical simulations predict that stars retain a metal content depending mainly on the total stellar mass of the system and that stellar metallicity remains practically static with time \citep{Torrey2019}. We will use Eq.~\ref{eq:met_model} to model the time evolution of stellar metallicity; even though this expression does not provide precise predictions for stellar metallicity, it sets an upper limit for the impact of variability in this property on galaxy colours. Assuming an equivalence between gas-phase and stellar metallicity may lead to a significant error in stellar metallicity. Nevertheless, this assumption has little impact on the results of this section because colour fluctuations induced by metallicity variations are largely insensitive to the actual stellar metallicity of a galaxy.

The dust content of a galaxy also reflects a litany of physical processes that are tightly connected to its star formation history. For example, supernova remnants and the envelopes of AGB stars have capacity to produce prodigious quantities of new dust \citep[e.g.,][]{Dwek1998, daCunha2010}, while shocks emanating from supernovae can destroy dust grains present in an enriched interstellar medium \citep{Calura2008, slavin_etal_2015}. Motivated by observations of the tight correlation between UV luminosity and dust attenuation \citep[e.g.,][]{Brinchmann2004}, we model the attenuation of interstellar light by dust with the same scaling relation used in \citet{behroozi_etal19},
\begin{eqnarray}
\label{eq:attenuation}
    \tau_{\rm UV}^{\rm ISM} = 1 + \mathrm{erf}\left(\frac{M_{1500}+20.93}{-3}\right),
\end{eqnarray}
where $\mathrm{erf}$ denotes the error function, while $M_{1500}$ refers to the absolute magnitude of an unattenuated galaxy at $1500\,\angstrom.$ To model the attenuation of galactic light by dust in this section, we start by fixing the dust attenuation in \fsps to zero and extract $M_{1500}$ using a $50\,\angstrom$-width top-hat filter. We then compute the components of our fiducial dust model by assuming that $\tau_\mathrm{V}^\mathrm{BC} = 3\,\tau_\mathrm{V}^\mathrm{ISM}$ (see \S\ref{sec:model_introduction}) and using that the dust attenuation curve evolves as a power law. 

Even though Eq.~\ref{eq:attenuation} does not capture the full complexity of physical processes controlling the time-variation of dust production and destruction, galaxy formation models predict a strong correlation between total dust formation rate and star formation rate \citep[e.g.,][]{triani2020_OriginDustgalaxies}, and thus between dust attenuation and star formation. Thus, despite our adopted model does not contain physical ingredients such as long-timescale dust production in AGB and planetary nebulae, our dust attenuation model serves the purpose of studying the impact of unresolved short-term variability in dust attenuation on galaxy colours. On the other hand, our model does not capture time variability in the dust attenuation curve caused by variations in the geometry of the star-dust mixture, which are likely to have a strong effect on UV rest-frame wavelengths \citep{Salim2020}.

The salient feature of these metallicity and dust attenuation models is that they have explicit dependence upon the {\em instantaneous} value of star formation rate. Thus by using these prescriptions, short-term SFH fluctuations will trigger short-term variability in metallicity and dust, allowing us to study how such variability influences the observed colours of a galaxy. In order to isolate the specific effect of this variability, we proceed as follows. For each snapshot at which SFHs are tabulated in \tng and \um, we compute the corresponding metallicity and dust attenuation using the above prescriptions together with the SFHs taken directly from the simulations, referring to the resulting time series as the exact metallicity and dust attenuation histories (ZHs and DAHs, respectively). We will furthermore repeat this calculation using the PCA-based approximate models for SFH, allowing us to generate approximate ZHs and DAHs with adjustable levels of short-term variability in metallicity and dust.

Fig.~\ref{fig:colours_gal_properties} gives a visual demonstration of the above models for a randomly selected \um galaxy; in the top, top-middle, bottom-middle, and bottom panels, we display the history of star formation, dust attenuation, metallicity, and stellar mass, respectively. Blue lines show results based on the exact \um SFH, while orange and green lines display results for histories based on PCA-based approximations using 5 and 20 components, respectively. As expected, we find that the metallicity and dust attenuation histories produced using the exact SFH in \um possess greater levels of short-term variability relative to those generated using PCA-based approximations. We can also see that the time evolution of dust attenuation and star formation is quite similar, reflecting the tight correlation between SFR and UV luminosity. Taken together with the fact that PCA-based models reproduce SFHs in an unbiased fashion (see \S\ref{sec:model_performance}), we do not expect systematic differences in DAHs to arise from PCA-based approximations to SFH. On the other hand, the difference between ZHs produced using the exact and approximate SFHs is more significant. We can understand how this effect arises in terms of the joint dependence of our metallicity model upon both stellar mass and SFR. First, SFH models resolving only long-term fluctuations result in somewhat inaccurate stellar mass histories, which can lead to an overall bias in ZHs. Additionally, we find that the amplitude of metallicity fluctuations is smaller than that of corresponding star formation fluctuations; this is likely due to the much stronger dependence of stellar metallicity on stellar mass relative to SFR \citep{Tremonti2004, ellison2008_CLUESORIGINMASSMETALLICITY, Mannucci2010}. While these results depend on the particularities of our metallicity model, the \tng simulation itself predicts metallicity variations exhibiting amplitudes and SFR-correlations that are quite comparable to those in our adopted model \citep{Torrey2019}.

We now use these models to examine how fluctuations in metallicity and dust attenuation manifest in the colours of galaxies. As we have already studied the impact of SFR fluctuations in the previous two sections, here will pinpoint the precise influence of fluctuations in metallicity and dust attenuation. For each SFH taken directly from the simulated catalogue, we calculate a ZH and a DAH as described above, and then we use these histories to calculate galaxy colours with \fsps. We also calculate galaxy colours using the ZHs and DAHs derived from the PCA-approximated models, while at the same time continuing to use the exact SFH from the simulation. Comparing the difference between colours produced using these two configurations allows us to investigate the specific influence on galaxy colours of physically-motivated levels of variability in metallicity and dust attenuation. By generating approximate ZHs and DAHs with increasing numbers of principal components, we are able to capture successively shorter-term fluctuations in metallicity and dust attenuation histories in an analogous fashion to how we previously studied variable levels of burstiness in star formation histories in \S\ref{sec:colours_burstiness}.

In Fig.~\ref{fig:colours_precision_properties}, we display the standard deviation of the difference between colours produced using exact vs approximate histories. Circles show results where the difference in colours arises only from the approximation of DAH, but using the exact SFH from the simulation and the ZH calculated from the exact SFH. Squares show the analogous results arising from the approximation of ZH, but with the exact SFH and corresponding exact DAH. Therefore, circles and squares isolate the influence on colours of short-term fluctuations in dust attenuation and metallicity, respectively. Calculations for colours observed through different filters are colour-coded as indicated in the legend. The axes in Fig.~\ref{fig:colours_precision_properties} are the same as those in Fig.~\ref{fig:colours_precision}: smaller values of the horizontal axis correspond to approximations generated with larger numbers of PCs, thereby capturing shorter-term fluctuations with greater precision. We remind the reader that the quantity $\sigmacol$ on the vertical axis indicates the standard deviation of the difference between colours produced using exact and approximated histories. The value of $\sigmacol$ increases monotonically as the x-axis increases, and thus the precision in colour decreases as the accuracy of the approximated histories degrades.

In comparing results for different coloured points of the same kind, we can see the impact of metallicity and dust attenuation variations is more substantial on bluer colours. For the case of metallicity, this is because newly-formed, bluer main sequence stars dominate the light of galaxies with lower metallicities \citep[e.g.,][]{Maraston2005}, while for the case of dust, this is because the level of attenuation is preferentially larger at shorter wavelength \citep[e.g.,][]{wang1996_InternalAbsorptionLuminosity}. Consequently, the impact of star formation, metallicity, and dust attenuation on photometry presents the same general trend as a function of wavelength.

The dashed lines in Fig.~\ref{fig:colours_precision_properties} indicate our best-fitting model to the impact of star formation variability on galaxy colours (see Fig.~\ref{fig:colours_precision}), and thus by comparing these lines to the circles and squares of the corresponding colour, we can assess the relative importance of star formation fluctuations in comparison to variability in dust attenuation and metallicity, respectively. As we can see, in all bands, the influence of variability in both dust attenuation and metallicity is much smaller than that of star formation fluctuations. Furthermore, we have verified that the collective impact of star formation, metallicity, and dust fluctuations is approximately consistent with adding in quadrature the impact of each property separately.

Our findings indicate that physically-motivated variations in stellar metallicity and dust attenuation present an essentially negligible influence on colours relative to that of physically-motivated star formation fluctuations. However, this result depends on the particularities of the metallicity and dust attenuation models: a stronger response of these properties to star formation fluctuations would lead to a more substantial impact of metallicity and dust attenuation variability on colours. The weak correlation between stellar metallicity and star formation rate, together with the reduced impact of metallicity variability on colours, suggests that the impact of metallicity variability on colours is smaller than that of burstiness even for extreme metallicity models. On the other hand, a model predicting much stronger dust production after star formation episodes could lead to a similar impact of dust attenuation and star formation variability on colours.


\section{Impact of variability on modelling galaxy colours}
\label{sec:forward}

In this section, we leverage the methodology introduced in \S\ref{sec:model} to study how short-term variability in star formation history impacts the modelling broad-band galaxy colours. In \S\ref{sec:forward_precision_colors} we focus on the colours of individual galaxies, and in \S\ref{sec:forward_scatter} we study the statistical distribution of colours of galaxy populations.


\begin{figure}
    \centering
    \includegraphics[width=\columnwidth]{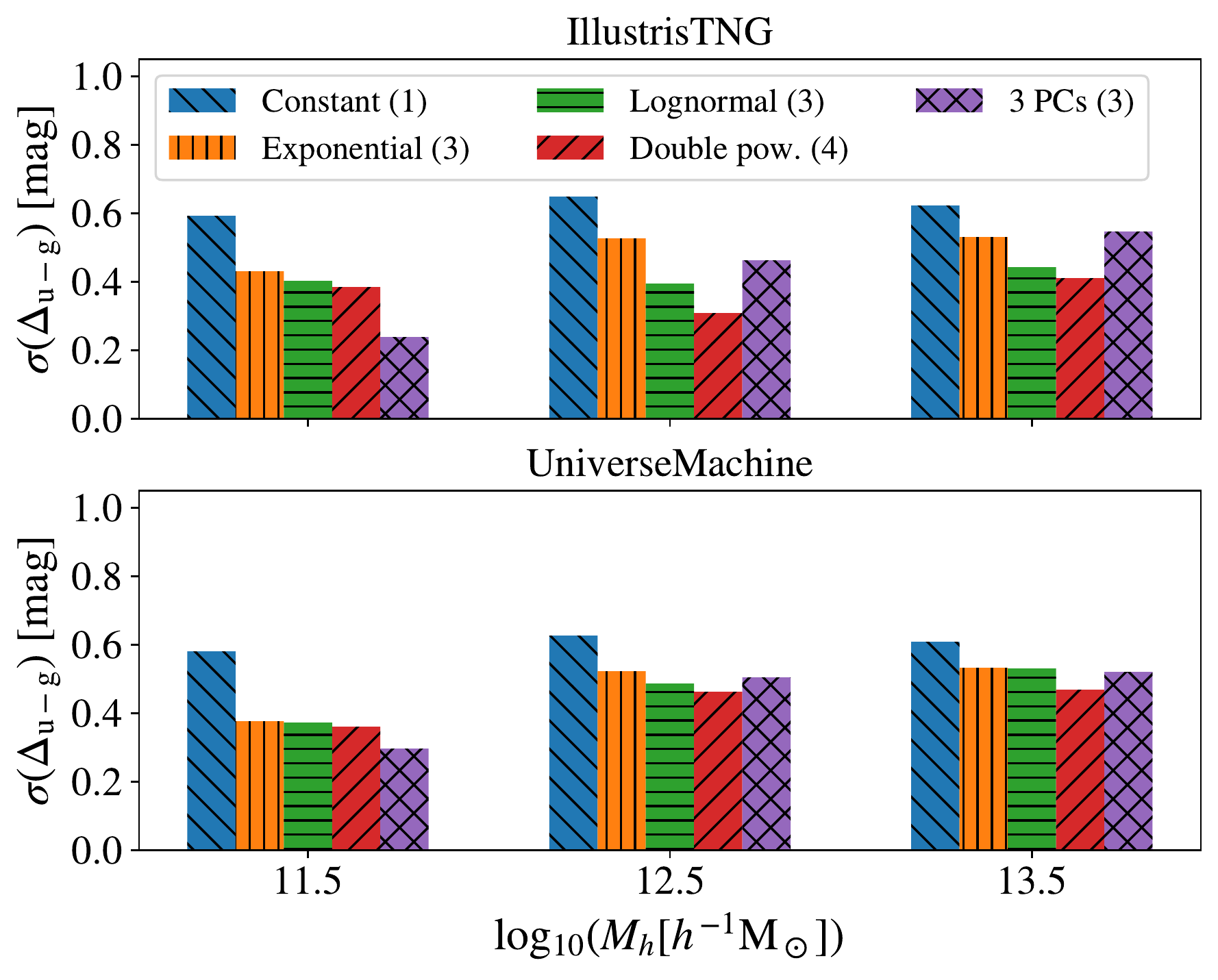}
    \caption{Precision of u-g colours produced using standard SFH models. The top and bottom panels display the results for different \tng and \um SFH samples; blue, orange, green, and red bars show the standard deviation of the difference between colours produced using original SFHs and best-fitting constant, delayed exponential, lognormal, and double power-law models, respectively; purple bars indicate the results for a 3-PC model. Parenthetical numbers next to the name of each model indicate its degrees of freedom. Overall, models with more degrees of freedom produce increasingly accurate colours, although each model's precision depends on its specifics and the target SFH sample. Strikingly, none of the models achieves precision better than 0.2 mag in the u-g colour for any sample.}
    \label{fig:forward_models}
\end{figure}

\begin{figure*}
    \centering
    \includegraphics[width=0.33\textwidth]{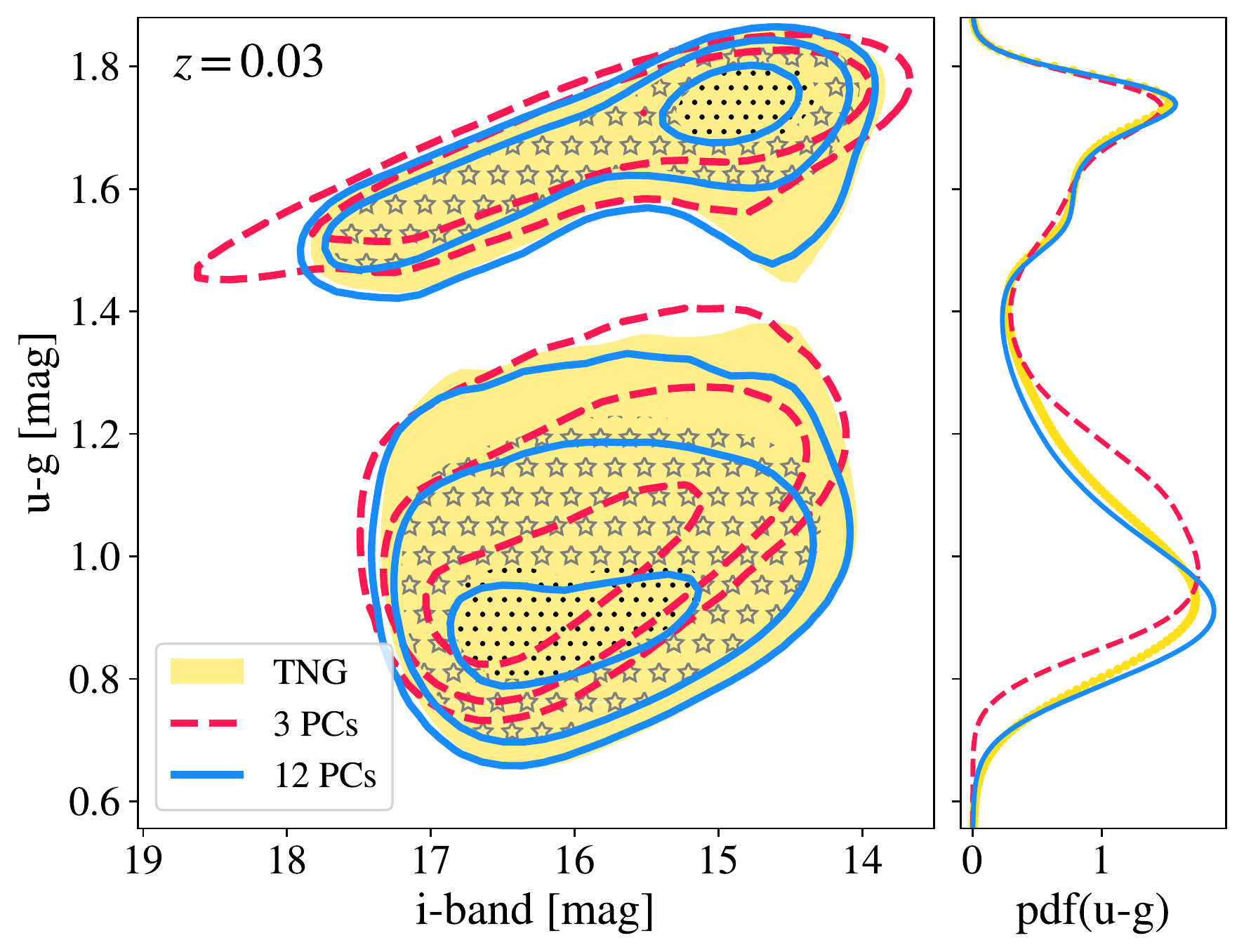}
    \includegraphics[width=0.33\textwidth]{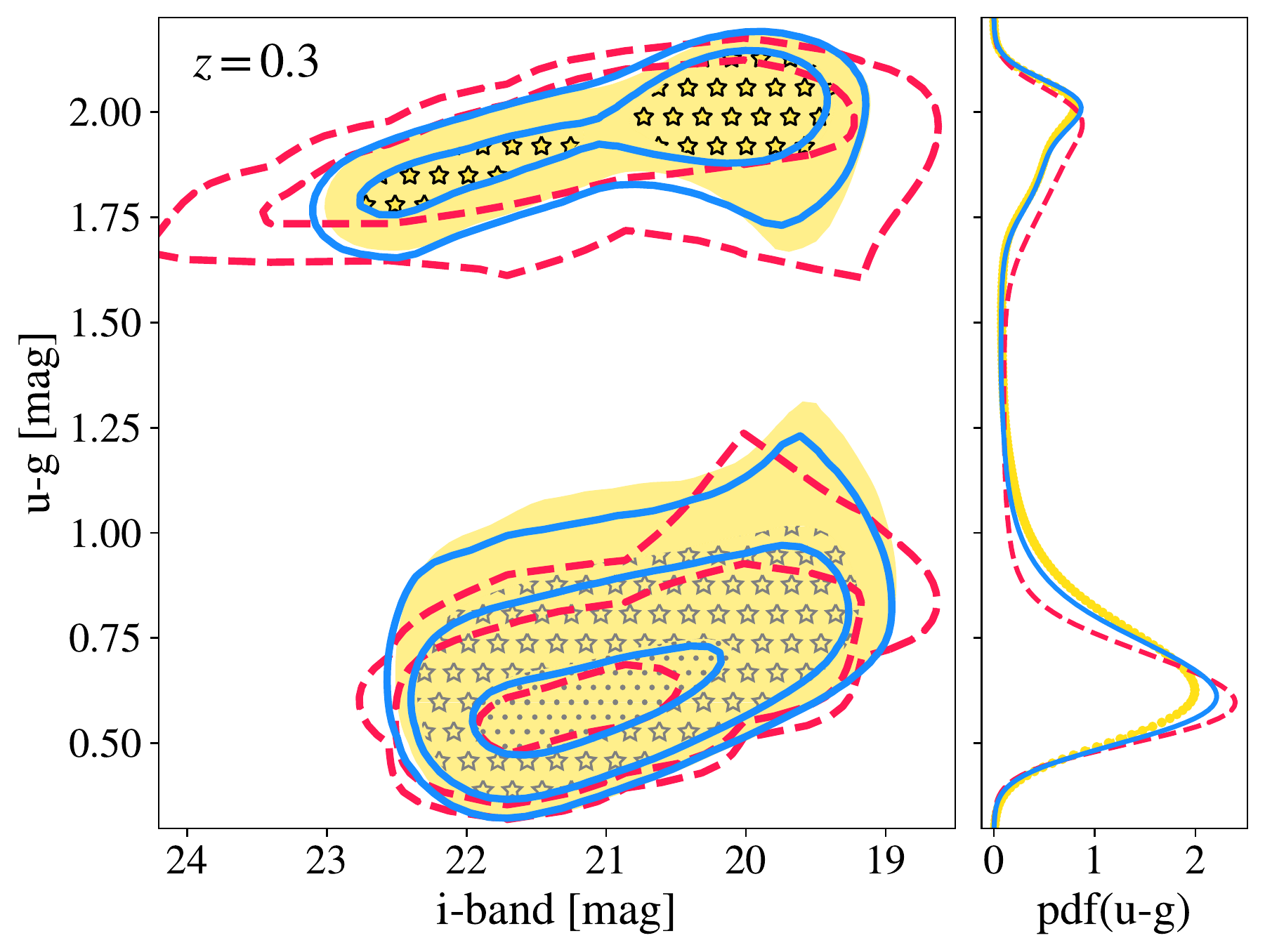}
    \includegraphics[width=0.33\textwidth]{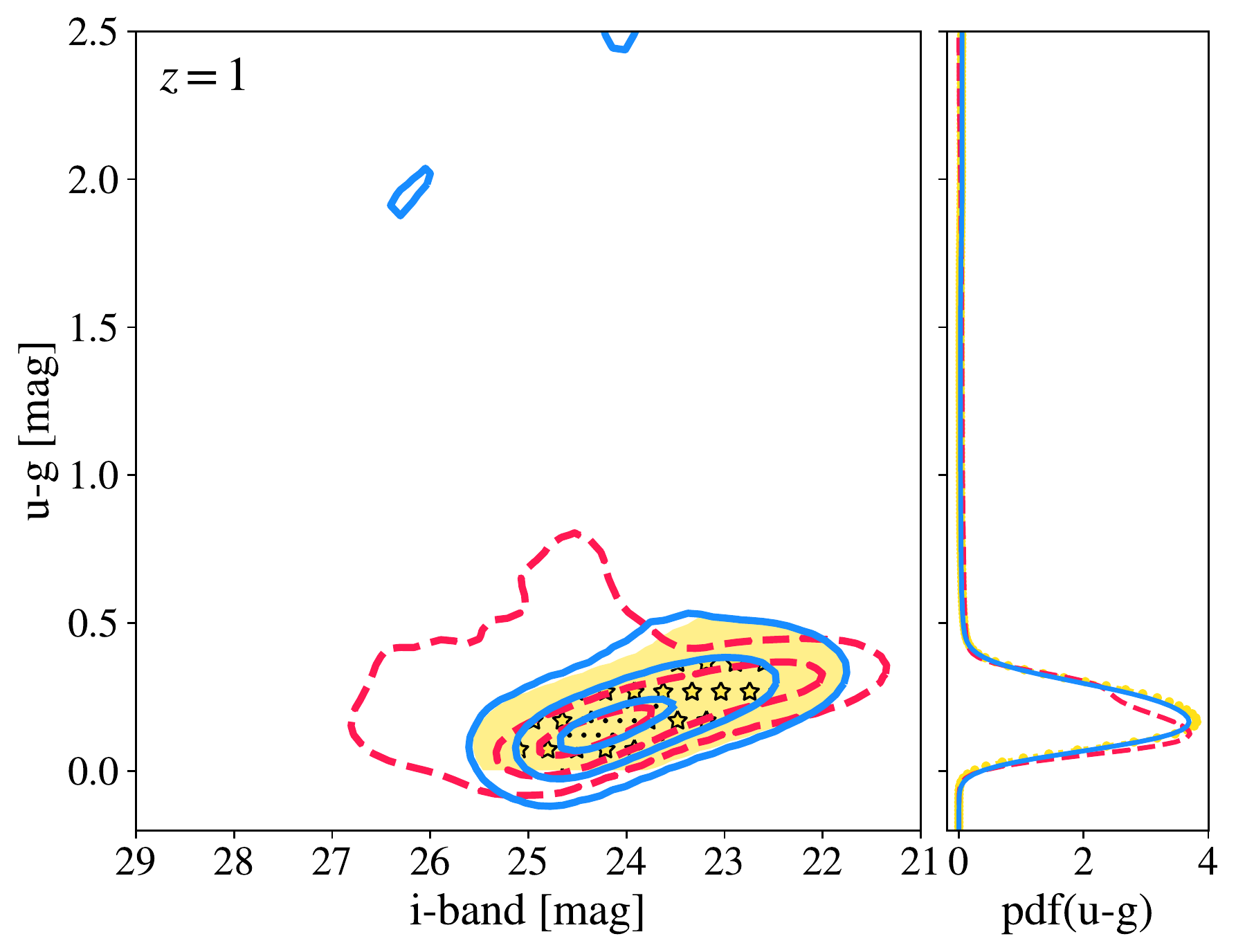}
    \caption{
    Distributions of colours produced using the SFHs of all \tng galaxies with $\Mstar>10^{9.25}\Msun$. Yellow shaded regions indicate colours generated using exact SFHs, while red dashed (solid blue) lines display colours produced using SFH approximations based on 3-PC (12-PC) models. The left, middle, and right panels show the results at $z=0.03$, 0.3, and 1, respectively; inner, intermediate, and outer levels enclose 25, 75, and 90\% of the galaxies. At each redshift, the larger left portion of the panel displays the 2-d distribution of u-g colour vs i-band observed magnitude, while the smaller right portion shows the marginalised distribution of u-g colour. The distribution of colours produced using exact SFHs is well-approximated by the distributions generated using both PCA-based approximations, indicating that the impact of variability on the colours of a galaxy population is minimal.}
    \label{fig:forward_scatter}
\end{figure*}

\subsection{Individual galaxies}
\label{sec:forward_precision_colors}

One of the most common approaches used to infer the physical properties of individual galaxies from their emitted light is to forward-model their SEDs using SPS models \citep[for a review of these models see][]{Conroy2013}. To do so, most investigations employ SFH models controlled by just a few degrees of freedom, such as exponentially decaying parametric forms and piecewise non-parametric formulations. In this section, we estimate the impact of unresolved burstiness on galaxy colours produced using some standard SFH models.

To assess the precision of galaxy colours resulting from a particular SFH model, we start by selecting a sample of SFHs for which we would like to produce colours, and then we compute the best-fitting solution of the target parametric model to each SFH in the sample. We continue by computing the logarithmic difference between each SFH and its best-fitting model, $\Delta_\mathrm{SFR}$. To quantify the level of variability that is not captured by the parametric model, we compute the standard deviation of this difference while considering all galaxies in the sample, $\sigmasfr$. Finally, we use our scaling relation calibrated using \fsps and SFHs predicted by \tng and \um to estimate the impact of star formation fluctuations on galaxy colours (see Eq.~\ref{eq:evar_col}). For our samples, we use SFHs predicted by \tng and \um for galaxies hosted by $\logMpeak\simeq11.5$, 12.5, and 13.5 haloes, and we fit these SFHs using each of the following four parametric models:
\begin{align*}
\dot{M}_*(t) = \left\{ \begin{array}{cc}
                A & {\rm constant} \\[0.5em]
                A(t-t_0)\exp[-(t-t_0)/\tau] & {\rm delayed\ exponential} \\[0.5em]
                (A/t) \exp[-(\log t - \log t_0)^2/\tau^2] & {\rm lognormal} \\[0.5em]
                A\left[ (t/\tau)^\alpha + (t/\tau)^{-\beta} \right] & {\rm double\ power\ law}
                \end{array} \right.
\end{align*}
where in the previous expressions, $A$, $t_0$, $\tau$, $\alpha$, and $\beta$ are the free quantities of the four parametric models.\footnote{Note that for the delayed exponential model, a clip is implemented such that $\dot{M}_*(t<t_0)=0.$}

In Fig.~\ref{fig:forward_models}, we show the precision of u-g colours produced using the SFH models mentioned above. The top and bottom panels display the results for different \tng and \um SFH samples, respectively. Blue, orange, green, and red bars show the standard deviation of the difference between colours produced using original SFHs and best-fitting constant, delayed exponential, lognormal, and double power-law forms, respectively; purple bars show the results for a PCA-based approximation based on 3 PCs. Each model's number of degrees of freedom is shown parenthetically next to its name in the legend. As we can see, models with more parameters generally produce increasingly precise colours, a simple reflection of their ability to leverage a greater number of degrees of freedom. Additionally, at a fixed number of parameters, models that better capture long-term SFH fluctuations produce more accurate colours; this can be seen in the performance of the lognormal model compared to that of the decaying exponential.

Other quantitative comparisons appear to depend on the specifics of the SFH model and the properties of the target SFH sample. For example, as shown in \citet{behroozi_etal19}, the {\em average} SFH of galaxies in \um exhibits double power-law type behaviour; consequently, the double power-law model shown in Fig.~\ref{fig:forward_models} is able to leverage its four degrees of freedom to achieve more accurate colours relative to the 3-PC model for galaxies hosted by $\logMpeak\simeq12.5$ and 13.5 haloes. However, galaxies hosted by $\logMpeak\simeq11.5$ haloes experience a greater variety of environmental influences that can produce wide variations in SFH, which may help understand the poorer performance of the 4-parameter double power-law model relative to the 3-PC model for these galaxies in both \tng and \um. While models with more degrees of freedom generally result in galaxy colours with increasing precision, each model's specific precision depends significantly upon its flexibility to capture physically-motivated SFHs. Results based on PCA may also be useful to estimate the level of systematic error incurred by other similarly formulated models. For instance, {\sc speculator} \citep{alsing2020_SPECULATOREmulatingstellar} models SFHs using a linear combination of four basis functions, and so this model should be expected to achieve a level of accuracy in galaxy colours comparable to a 4-PC model.

We also find that for all models and samples considered here, the imprecision of u-g colours is in excess of 0.2 mag. Thus in keeping with previous results, systematic errors associated with SFH models controlled by only a few free parameters should be expected to dominate the statistical error budget of present-day measurements. As explained in \S\ref{sec:colours}, these systematics arise from the limited ability of these models to capture short-term star formation variability. In principle, this conclusion depends upon the spectrum of star formation fluctuations predicted by \tng and \um, although we note that most state-of-the-art galaxy formation models predict a similar level of burstiness as these two models \citep{Iyer2020}. These findings imply that modelling the colours of individual galaxies with percent-level precision requires resorting to SFH models with more complexity than those considered in this section.

The results shown in Fig.~\ref{fig:forward_models} are based on the scaling relation between $\sigmacol$ and $\sigmasfr$ introduced in Eq.~\ref{eq:evar_col}, which we calibrated using PCA-based models. By design, these models describe SFHs in an unbiased fashion; therefore, SFH models not presenting this characteristic could lead to colour histories presenting a non-negligible bias relative to original colour histories. Using the SFHs of the 3\,015 \tng galaxies hosted by haloes of masses $12.45<\logMpeak<12.55$, we find that there is an average bias of $\left<\Delta_c\right>=0.80,$ 0.26, -0.01, -0.06, and -0.01 mag between the original u-g colour histories and those resulting from the best-fitting constant, delayed exponential, lognormal, double power law, and 3-PC models, respectively. Note that this bias presents a negligible impact on the predictions of Eq.~\ref{eq:evar_col}; we find that the average difference between the actual colour precision and that predicted by the model is within 10\%, and that model predictions do not degrade for SFH models presenting a larger bias.


\begin{figure}
    \centering
    \includegraphics[width=\columnwidth]{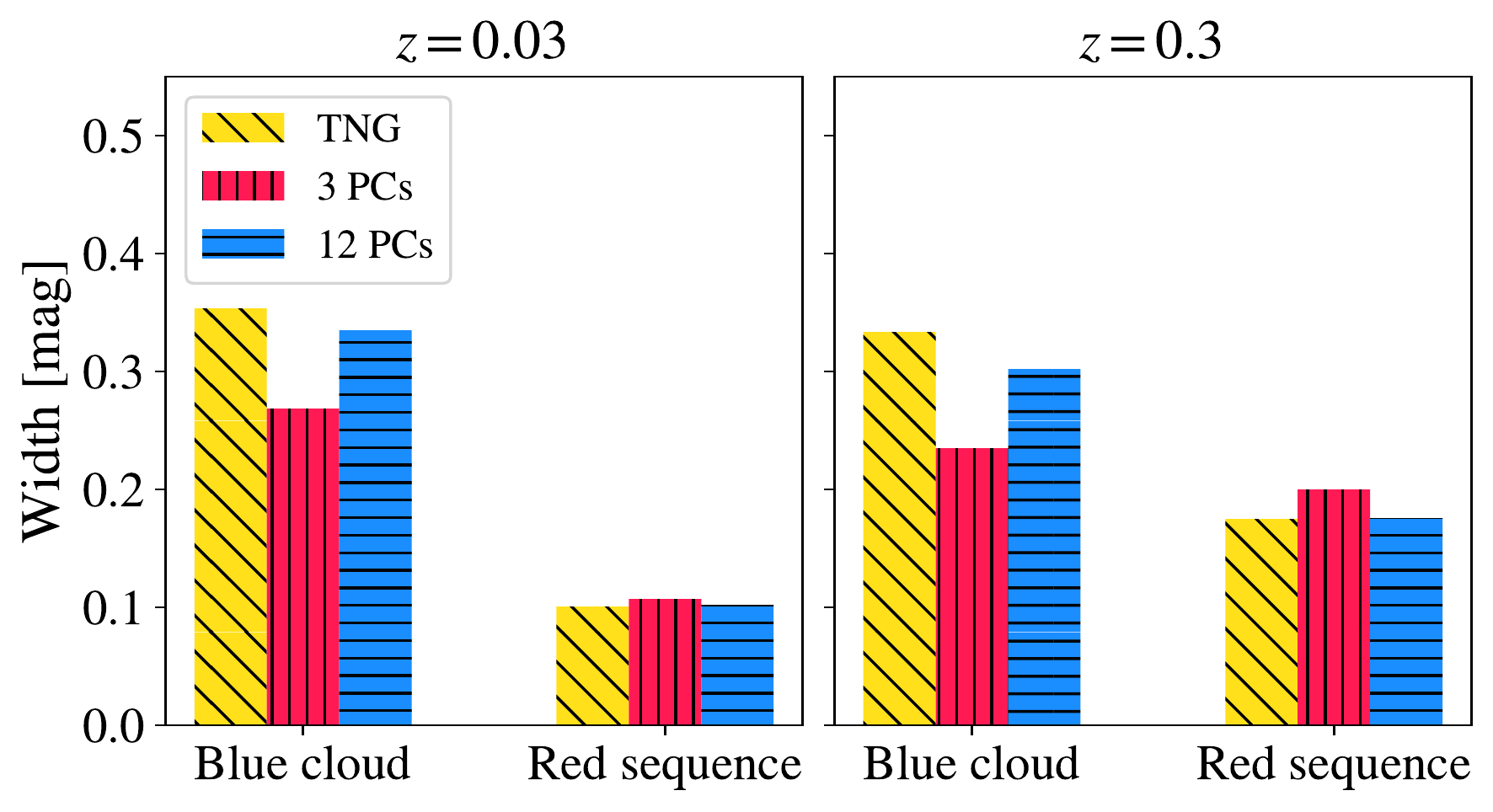}
    \caption{Impact of star formation variability on the widths of the blue cloud and red sequence (previously appearing in Figure~\ref{fig:forward_scatter}). The left and right panels display results for the width of the distributions in the u-g vs i-band plane at the redshifts shown at the top of each panel; yellow bars indicate results for colours produced using exact SFHs from \tng, while red (blue) bars denote results based on PCA-based approximations using 3 (12) components. SFHs capturing more variability result in colour distributions with a broader blue cloud and narrower red sequence.}
    \label{fig:forward_width_clouds}
\end{figure}

\subsection{Galaxy populations}
\label{sec:forward_scatter}

In the previous sections, we showed how the inaccuracy of SFH models resolving star formation fluctuations manifest in errors in the predictions for the colour of individual galaxies. We considered a variety of widely-used fitting formulae for SFH, as well as PCA-based models, and identified that the inability to resolve star formation variability is the dominant failure mode of all such models. In this section, we turn attention to a closely related question: how does the failure to resolve star formation variability translate into errors in the predictions for the colours of {\em populations} of galaxies?

To study the impact of variability on the distribution of colours for a galaxy population, we start by selecting the SFH of each \tng galaxy with stellar mass greater than $\Mstar=10^{9.25}\,\Msun$. For each SFH in the sample, we compute its PCA-based approximation using 3 and 12 PCs. Using exact SFHs and PCA-based approximations, we continue by computing metallicity and dust attenuation histories as outlined in \S\ref{sec:colours_properties}. Finally, we compute colours using exact and approximated histories. In the left, middle, and right panels of Fig.~\ref{fig:forward_scatter}, we display the resulting distribution of u-g colours and observed i-band magnitudes at $z=0.03$, 0.3, and 1, respectively. Yellow shaded regions, red dashed lines, and blue solid lines indicate the results for exact SFHs, 3-PC approximations, and 12-PC approximations, respectively; inner, intermediate, and outer contours enclose the 25, 75, and 90\% of the galaxies. As we can see, there is a close agreement between the colour distributions predicted by the exact and approximate SFH models; even colour-magnitude distribution resulting from the 3-PC model is a remarkably faithful representation of the exact distribution.

For a more quantitative analysis, we compute the width of the distribution of blue and red galaxies in Fig.~\ref{fig:forward_scatter}, which are commonly known as blue cloud and red sequence, respectively, using a two-component Gaussian Mixture Model. Note that the fraction of quench galaxies in \tng at $z=1$ is very small, explaining the absence of the red sequence in the right panel of Fig.~\ref{fig:forward_scatter}. In the left and right panels of Fig.~\ref{fig:forward_width_clouds}, we show the width of the blue cloud and red sequence across their minor direction at $z=0.03$ and 0.3, respectively. Yellow, red, and blue colours indicate the results for \tng SFHs, 3-PC approximations, and 12-PC approximations, respectively. The figure shows that the blue cloud's width increases for PCA-based approximations using more components, while the reverse is true for the red sequence: the width of the red sequence decreases as more PCs are included.

We can understand these trends in terms of the different physical processes that regulate these distributions. Bluer galaxies tend to be rapidly star-forming and bursty, and so as more PCs are included, more of this variability is accounted for, the breadth of the diversity of SFHs is better captured, and the width of the blue cloud increases. Meanwhile, galaxies in the red sequence are not bursty, but rather have experienced a rapid departure from the main sequence as they are quenched. As discussed in CH20 (see fig.~6 of CH20), this naturally produces a very tightly localised distribution of red colours \citep[see also][for a closely related discussion]{Matthee_Schaye2019}. Since PCA-based approximations are not physically informed by the rapid nature of galaxy quenching, this phenomenon can only be approximated through the linear combination of a sufficiently large number of components with different characteristic oscillation patterns (see the right-hand panel of Figure~\ref{fig:model_cartoon} for a visual reference). Therefore, as the number of PCs increases, the rapidity of quenching is better captured, and the width of the red sequence tightens to match the distribution predicted by \tng SFHs. In any case, the width of the blue cloud and red sequence varies by less than 0.1 and 0.025 mag across different samples; thus, even though short-term variability is missed entirely by the 3-PC model, this limitation plays only a minor role in the statistical distribution of broad-band galaxy colours. Consequently, even very coarse SFH models perform reasonably well in their ability to reproduce the width of the blue cloud and red sequence.

Beyond the agreement between the approximate and true colour distributions, we also find that star formation variability has a practically negligible impact on the decomposition of the galaxy population into star-forming and quenched based on their photometry. It is a widely-used practice to study red and blue galaxies separately due to their distinct internal properties and environmental characteristics. Based on the visually apparent bimodality in Fig.~\ref{fig:forward_scatter}, we divide the $z=0$ galaxy sample into subpopulations of blue and red galaxies according to the following cut: (u-g) $= 2.67 - 0.078\ m_i.$ Based on this criterion, we find that 31.8\% of galaxies are members of the red population according to colours produced using \tng SFHs, while the red fraction is 32.2 and 33.0\% for colours generated using the 3-PC and 12-PC models, respectively. Consequently, the division of galaxies into the red sequence and blue cloud is largely unaffected by physically realistic levels of SFH variability, especially in comparison to the influence of other factors such as dust obscuration \citep[e.g.,][]{maller_etal09, masters_etal10}.


\begin{figure}
    \centering
    \includegraphics[width=\columnwidth]{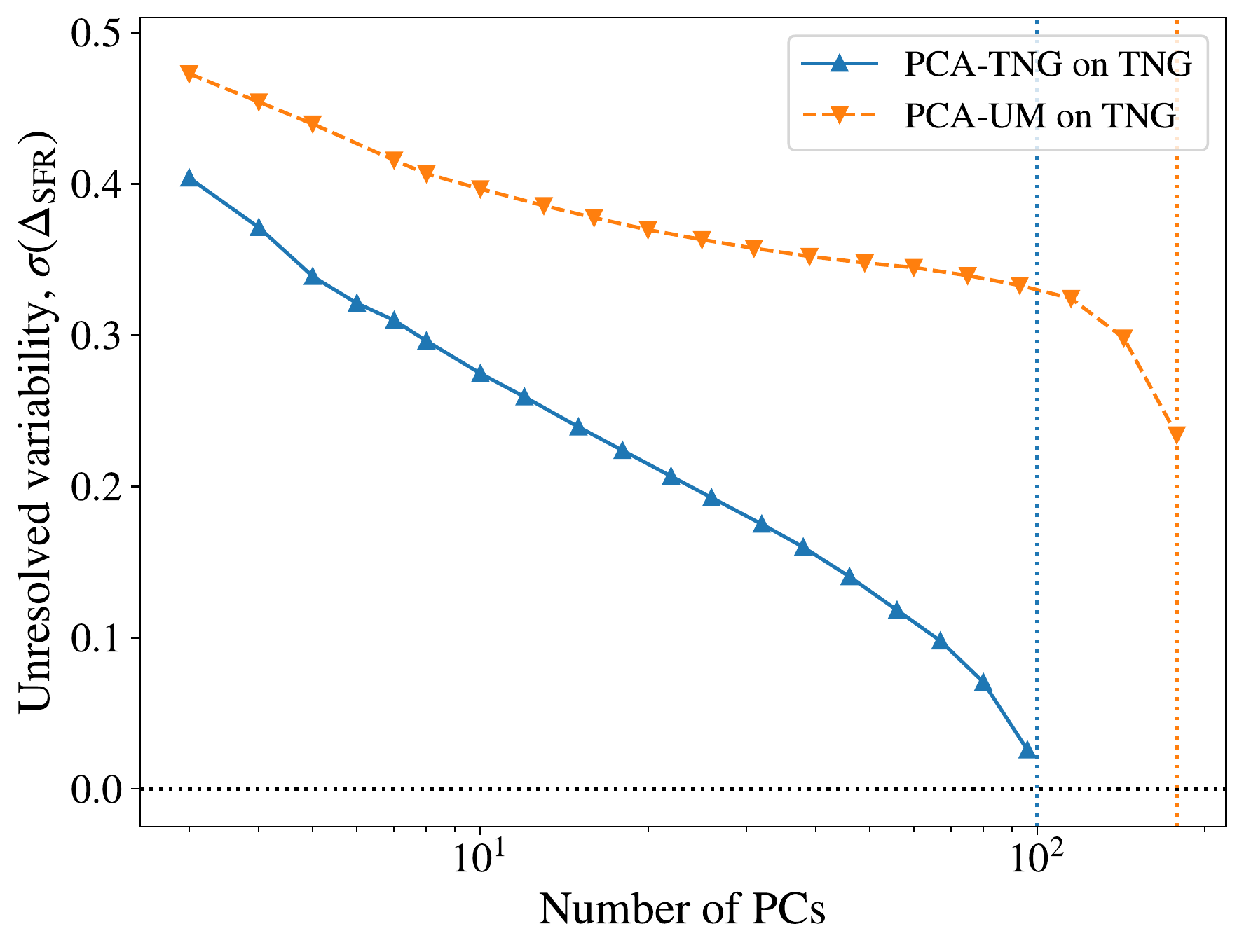}
    \caption{Limited generalizability of PCA-based approximations of SFH. Using galaxies in the \tng simulation as the target SFHs, the vertical axis shows the level of SFH variability not captured by the PCA model as a function of the number of PCs on the horizontal axis (see Eq.~\ref{eq:sigmasfr} for the definition of $\sigmasfr$). Each curve decreases monotonically because the success of any PCA-based approximation improves as the number of PCs increases. The blue curve shows the performance of a PCA model whose components were derived using the \tng simulation itself, while the orange curve shows the analogous results but instead using a model whose PCs were trained on the SFHs of galaxies in \um. We can readily see that the performance of the PCA model based on \um fares quite poorly in its ability to approximate the SFHs of galaxies in \tng.}
    \label{fig:discussion_generalize}
\end{figure}

\section{Discussion}
\label{sec:discussion}

This paper is the second in a series on Surrogate Modelling the Baryonic Universe, a new approach to forward modelling galaxy evolution in a cosmological context. Many of the basic precepts of SBU are shared with those of traditional semi-analytic models \citep[SAMs,][]{kauffmann_white_Guiderdoni_1993, baugh_cole_frenk_1996, avila_reese_1998}: the merger trees of dark matter halos are supplemented with a system of differential equations that regulates the galaxy formation physics that has been left out of the underlying N-body simulation \citep[see][for a comprehensive review]{somerville_dave_2015}. We highlight that in the SAM approach, the evolution of each quantity in a merger tree, evaluated at each simulated snapshot, is largely taken at face value to be the correct input to the system of equations governing the evolution of the galaxy; these equations are then solved on a halo-by-halo basis so that statistical scaling relations arise as emergent phenomena. By contrast, in the SBU approach, such statistical scaling relations are not purely emergent, but are treated as a foundational aspect of the model, a feature shared with contemporary semi-empirical models such as {\sc smad} \citep{becker_smad_2015}, {\sc emerge}, \citep{moster_etal18}, and \um \citep{behroozi_etal19}. However, rather than treating the evolution of each simulated halo exactly, as is done in SAMs and semi-empirical models, in SBU we rely on the existence of a smooth, parameterised family of solutions to many of the same differential equations used in SAMs, and we seek to identify minimally flexible mappings between the scaling-relation parameters and the properties of the underlying halos. Due to its fully parametric nature, our formulation admits an accurate, efficient, and highly scalable approximation with a differentiable surrogate model such as a neural network, as we will show in future papers in this series.

In the present paper, we have investigated a key question facing the SBU program: in approximating the star formation history of a galaxy, what level of complexity is required to capture the history of the galaxy's broad-band optical colour? We have studied this question using both conventional parametric forms of star-formation history, as well as a PCA-based approach, finding that models resolving shorter-term star formation variability result in increasingly precise colours. The fitting formula supplied by Eq.~\ref{eq:evar_col} provides a simple way to translate imprecision in SFH to an associated error in broad-band photometry, and applies to any filter centred at $\lambda_\mathrm{cen}$ with a typical broad-band width of a few hundred angstroms. While this scaling relation should be useful for such purposes, it neglects the fact that variations in colour induced by SFH variability are highly correlated. But the tightness of such correlations constitutes one of the principal findings of CH20: physically-motivated variations in SFH produce variations in only a single direction in colour-colour space, the SFH-direction, a result that is quite robust to the underlying galaxy formation model. As discussed in CH20, this result is encouraging for the prospect of forward modelling galaxy photometry, since it implies that the relationship between SFH and colour is relatively simple.

Fig.~\ref{fig:forward_scatter} of the present work provides further encouragement, as it demonstrates that the task of forward modelling the colours of {\em populations} of galaxies is significantly more straightforward than one would naively expect from attempting to predict the colours of individual galaxies. As discussed in \S\ref{sec:forward_scatter}, this result derives from the fact that galaxy colours depend primarily upon long-term star formation fluctuations, and short-term variability is largely decorrelated from such fluctuations. Figure~\ref{fig:forward_scatter} provides an explicit demonstration that an accurate forward model of the colours of galaxy populations does not necessarily need to capture the fine-grained details of the photometry of individual galaxies. This result serves as a specific example of the difference between the typical goals of the astronomer and the cosmologist, and the success of these efforts should be evaluated according to different metrics. The astronomer attempts to infer the properties of individual galaxies to uncover a detailed physical picture of the processes that have shaped individual systems and the galaxy population as a whole. On the other hand, the cosmologist is only concerned with the properties of individual galaxies insofar as such knowledge is required to make accurate predictions for the statistical distribution of galaxy properties across cosmic time. Moreover, in making such predictions, the distribution of galaxy properties is not directly the quantity of interest. Rather, the principal goal of the cosmologist is unbiased inference of cosmological parameters, and towards this end, the cosmologist needs only to understand the galaxy population with sufficient detail such that errors in predictions for the galaxy population are not degenerate with the cosmological parameters of interest. It is also worth pointing out that the extraction of cosmological constraints is evaluated in a fully Bayesian setting, so reducing the number of free parameters controlling the modelling of galaxy properties helps this endeavour.

Although we have shown that Principal Component Analysis provides a useful tool to relate SFH variability to galaxy colour, there are several features of PCA-based characterisation that make this technique poorly suited to achieve the goals of SBU. First and foremost, PCA-based approximations of SFH are essentially devoid of physical principles, and consequently, a PC decomposition trained on one simulation appears to generalise quite poorly when applied to another simulation. In Fig.~\ref{fig:discussion_generalize}, we show that a PCA-based model calibrated using the SFHs of \um galaxies shows much poorer performance in its ability to reproduce the SFHs of galaxies in \tng. The directly analogous result was shown in the appendix of \citet{Chen2020_empiricalSFH}, who found that a PCA-based model calibrated using \tng performs similarly poorly in its ability to reproduce SFHs predicted by the EAGLE simulation \citep{eagle_team_data_release_2017}. This result is daunting: in a true forward model of SFH, one needs to construct a mapping from each of the SFH parameters to dark matter halo properties, and the parameters of that mapping will need to be programmatically varied in an MCMC in order to account for uncertainty in the mapping reliably. Of course, such a mapping may be possible, but this result at least suggests that PCA-based models are more appropriate for creating a scaled-up replica of an existing hydrodynamical simulation, rather than as the basis of a forward model used in Bayesian analysis to interpret observational data.

When studying SFHs with standard parametric models, the assumed functional form effectively imposes unphysical prior expectations on the shape of star formation history, and it is now well known that this can lead to considerable biases on the inference of galaxy properties such as stellar mass, star formation rate, age, and dust attenuation \citep{simha_etal14, acquaviva2015_SIMULTANEOUSESTIMATIONPHOTOMETRIC, Pacifici2015, iyer_gawiser17, Carnall2019, Lower2020}. Non-parametric SFHs can mitigate such biases due to their greater degree of flexibility, enabling more robust constraints \citep{leja2018_OlderMoreQuiescent, leja_etal19, Lower2020, gilda2021_ScMirkwoodFast, thorne2020_DeepExtragalacticVIsible, nelson2021_SpatiallyResolvedStar}. However, this additional flexibility comes at a cost: the SFHs inferred using non-parametric methods are typically poorly constrained, a shortcoming that stems quite directly from the generality of these models. In fact, the posteriors on SFHs derived from non-parametric methods are commonly {\em prior-dominated}, and since high-resolution spectra with S/N$\gtrsim50$ are required to alleviate this issue \citep{leja_etal19}, then it appears that SFH inference based on non-parametric methods will continue to be prior-dominated when used to interpret cosmological survey data throughout the 2020s. Thus, in order to effectively harvest information about the physical characteristics of galaxies using non-parametric methods, the selection of appropriate prior information is of paramount importance. As shown in \citet{leja_etal19}, the determination of the form of these priors is highly nontrivial, and ultimately requires appeal to physical models of galaxy formation in order to validate the appropriateness of the selection.

The discussion of the role of priors in non-parametric methods provides a useful lens through which to view the shortcomings of PCA-based characterisations of SFH. By using a particular hydrodynamical simulation such as \tng to define a set of PCs, in effect this places a sharp prior on the physical model that underlies the simulation. Critically, there exists no unambiguous transformation of the PCs derived from one simulation to those of another; consequently, there is no clear way to incorporate systematic uncertainty associated with the choice of the simulation used to define the PC templates, even though such uncertainty is plausibly the {\em dominant} contribution to the error budget of the analysis.

Our future work on SBU is designed to address this situation with a complementary approach to the problem. In traditional parametric and non-parametric methods, the models are formulated in a physics-agnostic manner, and use a variety of physical simulations to validate their flexibility and assess robustness. In the SBU program, we instead use simulations to develop parametric functional forms that are based on the physics of galaxy formation, with parameters that directly encode physical processes such as dark matter halo growth, star formation efficiency, and quenching. Validation of the SBU functional forms proceeds in a similar fashion using a compilation of such simulations as well as observational data, which must necessarily include both simulations and datsets upon which the functional forms were not originally based. Therefore, the parametric formulation of SBU is specifically designed with flexibility to capture SFHs of distinct simulations; this formulation of the problem also offers a transparent means by which to derive SFH priors, since the parameters have a clear physical interpretation.

This effort is already well underway. The present paper has provided quantitative guidance on the specific requirements needed by an SFH model in order to achieve a target precision in predictions for broad-band colours. And in a companion paper to this work we present a differentiable model of the mass accretion history of dark matter haloes that will serve as the foundation of our parametric galaxy model. For purposes of predicting broad-band photometry, we will capitalise upon another key result of CH20: the influence of SFH upon colour is mainly determined by the fraction of stellar mass formed over the last $\tau_{\rm f}$ years before the time of observation, where $\tau_{\rm f}\approx1\,\mathrm{Gyr}$ contains the bulk of the information relevant for optical photometry, and most of the residual information is contained in $\tau_{\rm f}\approx0.25\,\mathrm{Gyr}.$ We will leverage this simplification in the next paper in this series to develop a forward model for those quantities, together with a differentiable surrogate.


\section{Conclusions}
\label{sec:conclusions}

Star formation variability is a generic prediction of all contemporary galaxy formation models, and reflects the different timescales of the physical processes regulating galaxy growth. In this work, we have studied the impact of this effect on modelling observed-frame broad-band optical and near-infrared colours for individual and populations of galaxies. To carry out our analysis, we leveraged the stellar population synthesis model \fsps, physically-motivated star formation histories predicted by the cosmological hydrodynamical simulation \tng and the empirical galaxy formation model \um, and SFH models with adjustable levels of variability based on principal component analysis. Our main findings can be summarised as follows:

\begin{itemize}
    \item We have provided a simple metric, $\sigmasfr,$ that quantifies the accuracy of an SFH approximation (see Eq.~\ref{eq:sigmasfr}). Using this simple metric, as well as a power spectrum analysis, we show that the \um galaxy formation model predicts significantly more burstiness relative to \tng (see Figs.~\ref{fig:model_power_spec} and \ref{fig:pca_expvar}).
    
    \item The precision of broad-band colours produced using SPS codes degrades as the level of unresolved burstiness increases, i.e., as the value of $\sigmasfr$ increases. The relationship between these two quantities can be faithfully captured with a simple fitting function (see Eq.~\ref{eq:evar_col}), which can be used to determine the minimal SFH model that leads to colours with target precision.
    
    \item Physically-motivated, short-term variability in metallicity and dust attenuation has a practically negligible impact on galaxy colours relative to burstiness (see Figure~\ref{fig:colours_precision_properties}).
    
    \item Modelling the colours of individual galaxies with percent-level precision demands resorting to complex SFH models, while producing precise colours for galaxy populations can be achieved using SFH models with just a few degrees of freedom (see Figs.~\ref{fig:forward_models} and \ref{fig:forward_scatter}).
    
    \item In dividing galaxies into quenched and star-forming subpopulations based on optical colours, the phenomenon of SFH burstiness has only a percent-level impact on the fraction of galaxies apportioned into each subpopulation, and so should be negligible in comparison to e.g., dust obscuration (see \S\ref{sec:forward_scatter}).
    \end{itemize}

Taken together, the previous findings encourage using parametric SFH approximations to forward modelling the colours of populations of galaxies. Having this motivation in mind, we introduce the next piece of SBU in a companion paper \citep{Hearin2021}: a differentiable model of the mass accretion history of haloes. The primary purpose of that model is serving as base for modelling galaxy SFHs, an approach that SBU shares with modern semi-empirical models such as {\sc smad} \citep{becker_smad_2015}, {\sc emerge}, \citep{moster_etal18}, and \um \citep{behroozi_etal19}. However, instead of treating the evolution of mass accretion histories from simulations directly as these models, in SBU we will rely on a parametric, differentiable model and we will seek to identify the minimal mapping between the scaling-relation parameters of the model and the properties of the underlying halos.


\section*{Acknowledgements}

We thank the anonymous referee for useful suggestions and comments. We acknowledge useful discussions with Raul Angulo. We gratefully acknowledge the use of the Atlas EDR cluster at the Donostia International Physics Center (DIPC), the Phoenix cluster at Argonne National Laboratory, which is jointly maintained by the Cosmological Physics and Advanced Computing (CPAC) group and by the Computing, Environment, and Life Sciences (CELS) Directorate, the Bebop cluster in the Laboratory Computing Resource Center at Argonne National Laboratory, and the facilities of the Yale Center for Research Computing. Work done at Argonne National Laboratory's work was supported by the U.S. Department of Energy, Office of Science, Office of Nuclear Physics, under contract DE-AC02-06CH11357.


\section*{Data availability}

We downloaded the \tng data used in this article from the IllustrisTNG data portal, \url{https://www.tng-project.org/data/}. We produced the \um data that we analyse by running the publicly available code\footnote{\url{https://bitbucket.org/pbehroozi/universemachine/}} on merger trees identified in the Bolshoi Planck simulation with Rockstar \citep{behroozi_etal13a} and ConsistentTrees \citep{behroozi_etal13b}. Our particular version of dataset will be shared upon request to the corresponding author.


\section*{Software}

This work made use of the following software packages: {\sc astropy} \citep{astropycollaboration2013_AstropyCommunityPython, astropycollaboration2018_AstropyProjectBuilding}, {\sc scikit-learn} \citep{pedregosa2011_ScikitlearnMachineLearning}, {\sc scipy} \citep{virtanen2020_SciPyFundamentalalgorithms}, {\sc fsps} \citep{Conroy2009, conroy_gunn_2020}, {\sc ipython} \citep{perez2007_IPythonSystemInteractive}, {\sc matplotlib} \citep{hunter2007_Matplotlib2DGraphics}, {\sc numpy} \citep{harris2020_ArrayProgrammingNumPy}, {\sc python-fsps} \citep{python_fsps}, and {\sc universemachine} \citep{behroozi_etal19}.


\bibliographystyle{mnras}
\bibliography{biblio}

\appendix
\renewcommand{\thefigure}{A\arabic{figure}}




\begin{figure}
    \centering
    \includegraphics[width=\columnwidth]{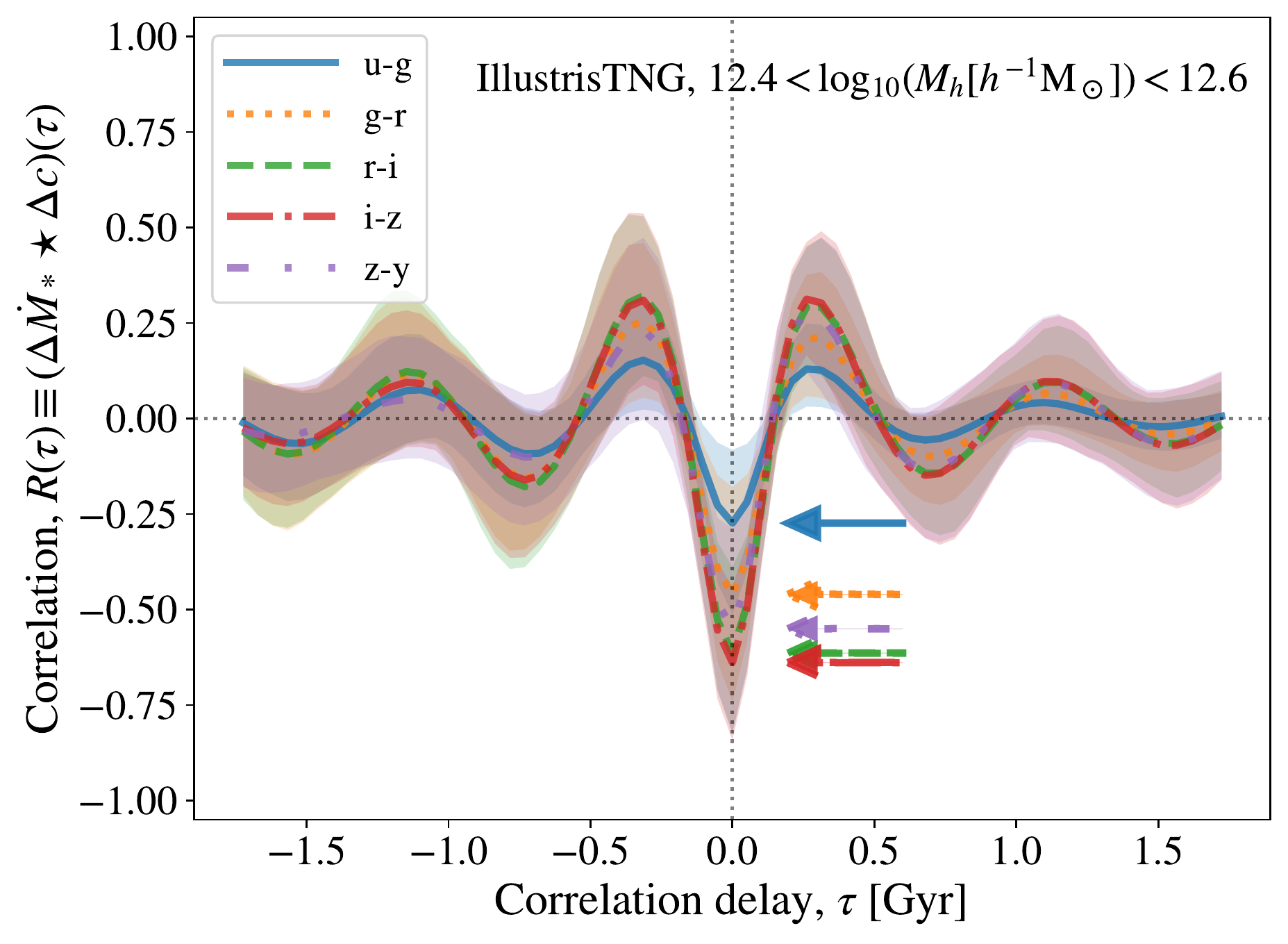}
    \caption{Time lagged cross-correlation between star formation and colour fluctuations extracted from \tng galaxies hosted by $\logMpeak\simeq12$ haloes. Colours indicate the results for distinct broad-band colours, lines and shaded areas denote averages and 16-to-84 percentile regions, respectively, and arrows point to the maximum absolute value of TLCCs. As we can see, the response of broad-band colours to star formation fluctuations is faster than $\tau\simeq0.14$ Gyr for all wavelengths, which is the shortest timescale resolved by publicly available \tng SFHs.}
    \label{fig:colours_time_lag_corr}
\end{figure}

\begin{figure}
    \centering
    \includegraphics[width=\columnwidth]{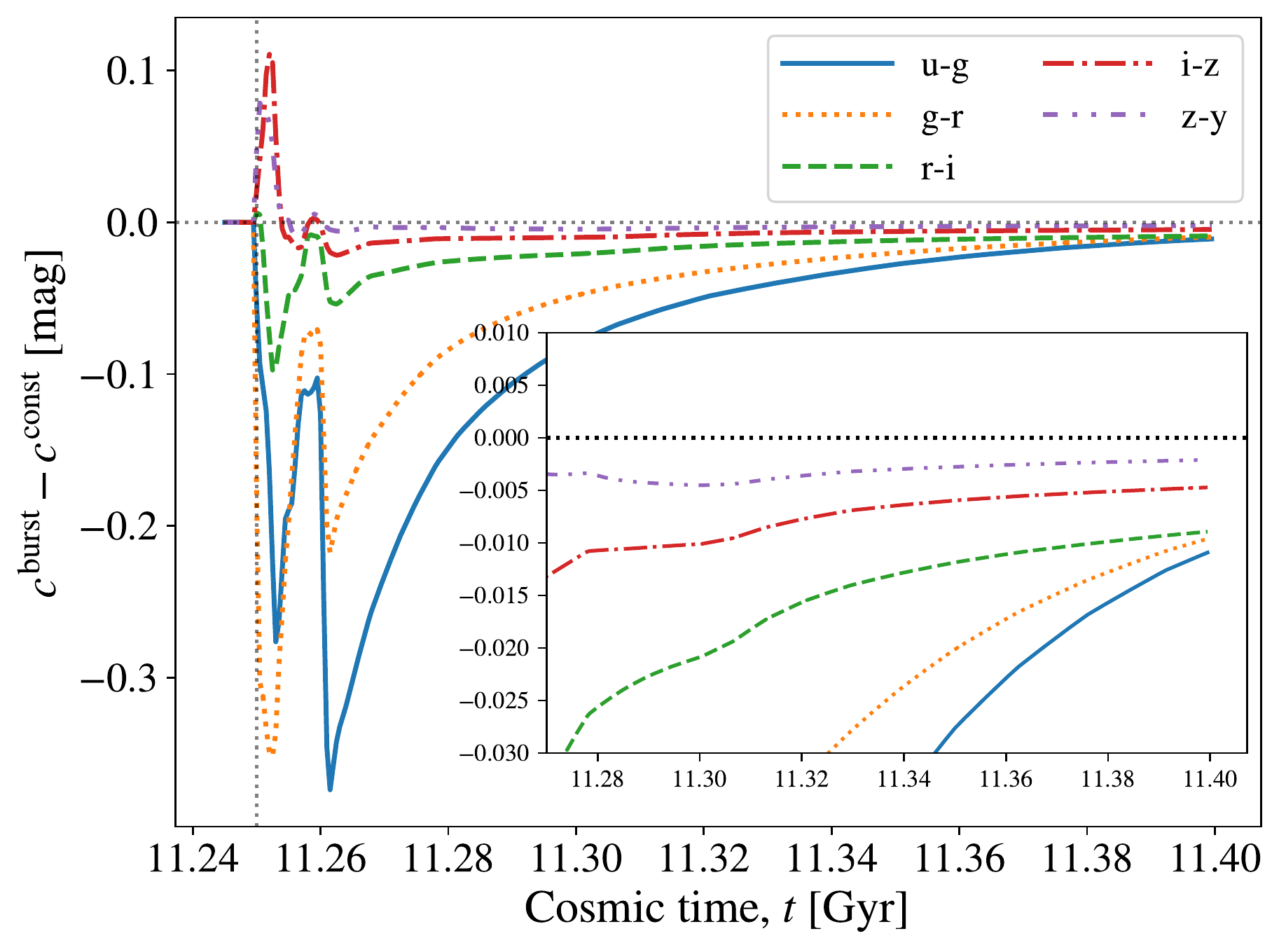}
    \caption{Response of broad-band colours to a star formation burst. Lines present the difference between colours produced using a constant and a constant plus burst SFH, colours indicate the results for different broad-band colours, and the inset displays the evolution of colours on long timescales. We find that galaxy colours present a monotonic response to star formation fluctuations on timescales longer than 0.01 Gyr, confirming that positive (negative) star formation fluctuations turn galaxy colours bluer (redder) on the timescales resolved by \tng and \um SFHs.}
    \label{fig:app_response}
\end{figure}

\section{Response of galaxy colours to star formation fluctuations}
\label{app:response}

In \S\ref{sec:colours_intro}, we noted an apparent correlation between star formation fluctuations and colour variations: positive and negative star formation fluctuations seem to turn galaxy colours bluer and redder, respectively. In this section, we use time-lagged cross-correlation (TLCC) to study whether the reported correlations are genuine.

Appealing to TLCCs is a standard strategy to analyse whether two simultaneous time series are correlated, and to determine the time shift for which such correlation is maximal. Before computing TLCCs, we first interpolate star formation and colour histories into 500 evenly-spaced cosmic times, and then we extract short-term fluctuations on these histories using PCA-based models; we interpolate histories to enable applying constant time shifts and we extract fluctuations to reduce spurious correlations arising from the smooth evolution of star formation and colour histories. Finally, we compute TLCCs as follows:
\begin{eqnarray}
    R(\tau) \equiv \left[\left(\dot{M}_*^\mathrm{orig} - \dot{M}_*^\mathrm{PCA}\right) \star \left(c^\mathrm{orig} - c^\mathrm{PCA}\right)\right](\tau),
\end{eqnarray}
where $\dot{M}_*^\mathrm{orig}$ and $\dot{M}_*^\mathrm{PCA}$ indicate exact SFHs and PCA-based approximations, respectively, $c^\mathrm{orig}$ and $c^\mathrm{PCA}$ denote colour histories produced using $\dot{M}_*^\mathrm{orig}$ and $\dot{M}_*^\mathrm{PCA}$, the $\star$ symbol refers to the cross-correlation operator, and $\tau$ denotes the time shift applied to star formation or colour histories when computing correlations.

In Fig.~\ref{fig:colours_time_lag_corr}, we display the TLCC of \tng star formation and colour histories for galaxies hosted by $\logMpeak\simeq12$ haloes. Colours indicate the results for distinct broad-band colours, lines and shaded areas denote averages and 16-to-84 percentile regions, respectively, and arrows point to the maximum absolute correlation. As we can see, TLCCs reach their absolute extreme at $\tau=0$, remain near their peak strength for $|\tau|<0.2$ Gyr, and drop rapidly to statistically insignificant values for longer time shifts. Thus, the response of broad-band colours to star formation fluctuations is faster than $\tau\simeq0.14$ Gyr for all wavelengths, which is approximately the shortest timescale resolved by publicly available \tng SFHs. We use PCA-based models presenting 55 PCs to extract fluctuations; we check that the results remain qualitatively unchanged when using models presenting a different number of PCs.

One of the most prominent features in this figure is that $R(\tau=0)$ is negative for all broad-band colours, confirming that positive and negative star formation fluctuations turn galaxy colours bluer and redder on timescales equal or longer than $\tau\simeq0.14$ Gyr, respectively. Nonetheless, it is worth noticing that this timescale is too long to resolve physical processes associated with massive stars, and thus it is conceivable that our findings may not hold for shorter timescales. To study the response of galaxy colours to star formation fluctuations on even shorter timescales, we compute colour histories with 0.5 Myr resolution using a constant SFH and a constant plus burst SFH. Specifically, we include a burst taking place at 11.25 Gyr, lasting for 0.5 Myr, and supposing an increase of two orders of magnitude in SFR. Fig.~\ref{fig:app_response} shows the difference between colours produced using these two SFHs. As we can see, the colours of the galaxy with a burst present rapid fluctuations relative to the colours of the galaxy with no burst over the first 10 Myr after the burst, reflecting the influence on galactic light of recently formed supergiants \citep{Charlot1991}. However, it is essential to note that SPS codes exhibit significant scatter in this regime \citep{conroy_etal10a}.

For timescales more extended than 0.01 Gyr after the burst, the galaxy with a burst presents bluer colours than the galaxy with no burst because the first forms a larger number of massive main-sequence stars than the second, and these stars dominate galactic light at short wavelengths. For even longer timescales, the colours of both galaxies asymptotically converge as massive stars formed during the burst depart from the main sequence. Taken together, these findings let us conclude that galaxy colours present a monotonic response to star formation fluctuations on timescales longer than 0.01 Gyr. This timescale is much shorter than that of \tng and \um SFHs, and thus it is safe to assume that star formation fluctuations result in instantaneous, monotonic variations in galaxy colours produced using SFHs drawn from these two models.

\bsp
\label{lastpage}
\end{document}